\documentclass[floatfix,twocolumn]{aastex631}
\usepackage{amsmath}
\usepackage[T1]{fontenc}
\usepackage{textcomp}
\usepackage{gensymb}

\newcommand{\rvar}{$R_{var}$}
\newcommand{\starspot}{\texttt{starspot}}
\newcommand{\celerite}{\texttt{celerite2}}
\newcommand{\lightkurve}{\texttt{lightkurve}}
\newcommand{\thorosmos}{\texttt{thorosmos}}

\newcommand{\stella}{\texttt{stella}}
\newcommand{\astropy}{\texttt{astropy}}
\newcommand{\ernlib}{\texttt{ernlib}}
\newcommand{\IRAF}{\texttt{IRAF}}
\newcommand{\thorsky}{\texttt{thorsky}}
\newcommand{\readmultispec}{\texttt{readmultispec}}
\newcommand{\pymc}{\texttt{PyMC3}}

\newcommand{\lhalbol}{$L_{H\alpha}/L_{bol}$}
\newcommand{\ha}{$H\alpha $}

\begin{document}

\title{Contemporaneous Observations of \texorpdfstring{\ha} \ Luminosities and Photometric Amplitudes for M Dwarfs}

\author[0000-0001-9828-3229]{Aylin Garc{\'i}a Soto}
\affiliation{Department of Physics and Astronomy, Dartmouth College, Hanover NH 03755, USA}

\author[0000-0003-4150-841X]{Elisabeth R. Newton}
\affiliation{Department of Physics and Astronomy, Dartmouth College, Hanover NH 03755, USA}

\author[0000-0001-7371-2832]{Stephanie T. Douglas}
\affiliation{Department of Physics, Lafayette College, Easton PA 18042, USA}

\author[0000-0002-5922-4469]{Abigail Burrows}
\affiliation{Department of Physics and Astronomy, Dartmouth College, Hanover NH 03755, USA}

\author[0000-0002-3239-5989]{Aurora Y. Kesseli}
\affiliation{Infrared Processing and Analysis Center, Caltech, Pasadena CA 91125, USA}



\begin{abstract}
While many M dwarfs are known to have strong magnetic fields and high levels of magnetic activity, we are still unsure about the properties of their starspots and the origin of their magnetic dynamos. Both starspots and chromospheric heating are generated by the surface magnetic field; they produce photometric variability and $H\alpha$ emission, respectively. Connecting brightness variations to magnetic activity therefore provides a means to examine M dwarf magnetism. We survey 30 M dwarfs previously identified as fast rotating stars ($P_{rot} < 10$ days). We present time-series optical photometry from the Transiting Exoplanet Survey Satellite (TESS) and contemporaneous optical spectra obtained using the Ohio State Multi-Object Spectrograph (OSMOS) on the 2.4m Hiltner telescope at MDM Observatory in Arizona. We measure rotation periods and photometric amplitudes from TESS light curves using Gaussian Processes. From the OSMOS spectra, we calculate the equivalent width of $H\alpha$, and \lhalbol. We find a weak positive correlation between $H\alpha$ luminosity and the semi-amplitude, \rvar \ ($p=0.005_{-0.005}^{+0.075}$). We also observe short-term variability (between 20-45 minutes) in $H\alpha$ equivalent widths and possible enhancement from flares consistent to recent literature.
\end{abstract}



\section{Introduction}\label{sec:intro}

M dwarfs make up $\sim$ 70\% of stars in our Galaxy \citep{henry_solar_2006}. Due to their smaller sizes, they produce deeper exoplanet transits and larger radial velocity amplitudes. Thus, they present an easier way to find Earth-size planets compared to FGK stars. 
However, M dwarfs also pose a problem because they are very magnetically active stars. Active regions on the stellar surface can produce large-amplitude radial velocity signals, masking or masquerading as planets \citep[e.g.,][]{boisse_disentangling_2011,haywood_planets_2014,newton_impact_2016,robertson_persistent_2020}. They also can produce residual features in transmission spectroscopy, masking or masquerading as planetary spectral features, especially water \citep[e.g.,][]{pont_detection_2008, rackham_transit_2018, zhang_near-infrared_2018, wakeford_disentangling_2019}. 
An in-depth understanding of magnetic activity in M dwarfs--starspots, activity properties, and magnetic dynamos--will support exoplanet research in the decades to come. 

Starspots on M dwarfs are assumed to form the same way as sunspots, with dark regions resulting from magnetic field lines rising to the stellar surface. In the Sun, the origin of the magnetic field is thought to be the $\alpha \Omega$ dynamo, where differential rotation and the tachocline, the boundary between the radiative zone and the convective zone, play an important role \citep{parker_hydromagnetic_1955, charbonneau_dynamo_2010,cameron_global_2017}. The properties of starspots are therefore an important observational signature of the magnetic dynamo \citep{berdyugina_starspots_2005}. 

Strong, large scale magnetic fields of M dwarfs have been detected using Zeeman broadening \citep{robinson_magnetic_1980, saar_improved_1988, valenti_infrared_1995} and Zeeman Doppler Imaging \citep{donati_large-scale_2008, morin_large-scale_2010}. However, M dwarfs of masses less than 0.35 $M_\odot$ are fully convective; they lack a radiative zone and consequently a tachocline \citep{chabrier_structure_1997}. In these stars, $\alpha \Omega$ is likely not the underlying dynamo generating the observed large-scale fields, and it is reasonable to expect different magnetic properties on M dwarfs compared to solar-type stars. 

The dynamo and magnetic fields can also be studied through the relationship between rotation period and activity indicators such as chromospheric emission (particularly Ca II H\&K and 
$H\alpha$) 
and coronal (X-ray) emission.  Chromospheric and coronal heating and the resulting emission is thought to reflect the underlying magnetic dynamo mechanism \citep{parker_hydromagnetic_1955,kim_theoretical_1996}.

Those magnetic field proxies are plotted against the Rossby number, the ratio of rotation period to convective overturn time. These relations show two regimes. In the unsaturated regime, activity increases with decreasing period or Rossby number. In the saturated regime, most studies show a plateau inactivity for rapid rotators, at a critical Rossby number $\lesssim 0.1$ \citep[e.g.,][]{reiners_evidence_2009}. This is seen in Ca II H\&K \citep[e.g.,][]{vaughan_comparison_1980, middelkoop_magnetic_1981, browning_rotation_2010}, $H\alpha$ \citep[e.g.,][]{mekkaden_rotation_1985, delfosse_rotation_1998, mohanty_rotation_2003, reiners_catalog_2012, douglas_factory_2014, newton_h_2017, nunez_rotation-activity_2017}, X-ray \citep[e.g.,][]{wright_stellar-activity-rotation_2011,nunez_rotation-activity_2017,wright_stellar_2018,magaudda_relation_2020}, magnetic field strength \citep[e.g.,]{donati_large-scale_2008,reiners_evidence_2009,vidotto_stellar_2014,shulyak_strong_2017,shulyak_magnetic_2019,reiners_magnetism_2022} and flares \citep[e.g.,][]{davenport_kepler_2016,medina_flare_2020,medina_galactic_2022}. However, others studies show a weak correlation between the Rossby number and activity index in the saturated regime \citep[e.g.,][]{mamajek_improved_2008, reiners_generalized_2014, lehtinen_knee_2021, magaudda_relation_2020}. 

The rotation-activity relations show scatter that appears to be intrinsic \citep{newton_h_2017, boudreaux_ca_2022}. \citet{newton_h_2017} noted a relationship between photometric variability amplitude, which derives from starspot-related modulations, and the $H\alpha$ emission line strength, measured relative to the bolometric luminosity, \lhalbol. They find a positive correlation between semi-amplitude and \lhalbol, and concluded that non-sinusoidal rotational variability and spot evolution were likely impacting the results by creating intrinsic scatter. 

This paper explores both potential sources of scatter in photometric amplitude, \rvar \, and \lhalbol. We address non-sinusoidal rotation by using uniformly-obtained, high S/N light curves and measuring amplitude non-parametrically. We address spot evolution through the use of contemporaneous photometric and spectroscopic observations. To do this, we use photometry from Transiting Exoplanet Survey Satellite \citep[TESS;][]{ricker_transiting_2015} and new spectroscopic data from the Hiltner Telescope 2.4m Ohio State Multi-Object Spectrograph \citep[OSMOS;][]{martini_ohio_2011}. 

In \S \ref{sec:data} we present our M dwarf sample and stellar parameters \S \ref{subsec:param}, in \S \ref{sec:obs} the instruments that we use, in \S \ref{sec:analysis} we explain the process of removing stellar flares, the analysis of the photometric and spectroscopic data, and the computation of stellar parameters. In \S \ref{sec:results} and \S \ref{sec:discussion} we compare our results with \citet{newton_h_2017} and lastly, conclude in \S \ref{sec:conclude}.

\section{Sample} \label{sec:data}
\subsection{Northern Hemisphere M dwarfs with rotation and \texorpdfstring{$v\sin i$} \ Measurements}\label{subsec:catalogs}

We cross-match nearby M dwarfs from the MEarth Project \citep{newton_rotation_2016} with known rotation periods against catalogs containing $v \sin i$ measurements for M dwarfs. We consider three $v \sin i$ catalogs that have large samples of nearby, mid- and late-type M dwarfs ($M < 0.5 M_{\odot}$) and which reliably distinguish between detections and non-detections. 
We then select stars in the Northern Hemisphere with $V < 17.5$ and rotational periods $P_\textrm{rot}\ <\ 13$~days (corresponding to half of a TESS sector) from the following catalogs: 

\begin{itemize}
    \item  \citet{fouque_spirou_2018} presents high resolution spectra of 440 M dwarfs (M0V-M6.5V) taken with the {\'E}chelle SpectroPolarimetric Device for the Observation of Stars \citep{manset_espadons_2003} at Canada-France-Hawaii 3.6m Telescope. This catalog is provides $v\sin i$ measurements for 5/30 M dwarfs in our final sample. 
    \item \citet{reiners_carmenes_2018} analyze high resolution spectra of 324 M dwarfs (M0V-M9V) using the Calar Alto high-Resolution search for M dwarfs with Exoearths with Near-infrared and optical {\'E}chelle Spectrographs \citep{quirrenbach_carmenes_2012} on the 3.5m telescope at the Calar Alto Observatory. FThis catalog is provides $v\sin i$ measurements for 9/30 M dwarfs in our final sample.  
    \item  \citet{kesseli_magnetic_2018} reports the $v \sin i$ of 88 rapidly rotating M dwarfs (M3V-M6V) with spectra taken from the Immersion GRating INfrared Spectrograph \citep{park_design_2014} on the Discovery Channel 4.3m Telescope and at the Harlan J. Smith 2.7m Telescope, as well as the iSHELL {\'E}chelle spectrograph \citep{rayner_ishell_2016} on NASA's 3.0m Infrared Telescope Facility. This paper specifically targets M dwarfs from the MEarth Project and so provides significant overlap with our list of rotating M dwarfs. This catalog is provides $v\sin i$ measurements for 16/30 M dwarfs in our final sample.
\end{itemize}

The merged catalog amounts to 133 stars that meet our brightness and $P_\textrm{rot}$ limits. Further cuts, as discussed below, reduce our target list to 30 stars.

\subsection{Identifying Unresolved Binaries}

Initial vetting for unresolved multiple systems was done in \citet{newton_h_2017} and the three  $v \sin i$ catalogs. We further query objects in our sample using the {\it Gaia} DR2 and DR3 catalog \citep[] []{gaia_collaboration_gaia_2016,gaia_collaboration_gaia_2018,gaia_collaboration_gaia_2022} for the Re-normalized Unit Weight Error (RUWE). RUWE is a chi-square metric of fitting models to single stars, which is intended to remove the dependency of color and magnitude of the {\it Gaia} Unit Weight Error \citep[][]{lindegren_re-normalising_2018,lindegren_gaia_2021}. Large astrometric errors can indicate binarity, with RUWE$\gtrsim1.2$ often used to identify likely binaries \citep{rizzuto_zodiacal_2018,bryson_occurrence_2021}. M dwarfs have a significantly lower binary fraction ($30\%$) than higher mass stars \citep{cortes-contreras_carmenes_2017,winters_solar_2019}, yet we find a large fraction of our sample have RUWE$>1.2$. 

\citet{winters_volume-complete_2021} find that the stars independently identified as binaries have large values of astrometric excess noise, distinct from the single star population. As discussed in \S 3.3 of that work and references therein, the majority of their sample are assessed for binarity using high-contrast imaging and/or high-resolution spectroscopy. Unsurprisingly, we find that RUWE is also effective at distinguishing the known binaries. However, the median RUWE for likely single, mid-to-late M dwarfs is RUWE$=1.2$, with the distribution extending to RUWE$=2.1$. 5\% of the likely single stars have RUWE$>1.6$, while none have RUWE$<0.8$. Thus, a different cut-off is needed to identify likely binaries in our sample. We evaluate the RUWE distribution for nearby, mid-to-late M dwarfs in the 15 pc volume-complete sample from \citet{winters_volume-complete_2021}.

We fit a Gaussian to the distribution of RUWE for the M dwarfs from \citet{winters_volume-complete_2021} expected to be single stars. Since the RUWE distribution is slightly skewed towards RUWE$>1.2$, we only fit the half of the sample at RUWE$<1.2$. The standard deviation ($\sigma_{std}$) of the fitted Gaussian is $0.13$. We adopt $3\sigma_{std}$, or RUWE=$1.6$ as the cut-off to identify likely binaries. 

We additionally check SIMBAD for references flagging the stars in our sample as binaries. For the stars with RUWE$>1.6$, we find the majority of stars were independently identified as binaries. We also manually remove 11 stars with RUWE$<1.6$ that have previously been flagged as close binaries. 
We retain three known binaries which are either more widely separated ($> 2170$ AU which is resolvable in Gaia) or with a dim $V>17.5$ companion, since we would not expect these companions to affect our results for the primary.

Lastly, we look at the TESS contamination ratios for these stars. This ratio is the fractional contribution of nearby stars divided by the flux of the target determined in \citet{stassun_tess_2018}. For this study, we limit the contamination ratio to $<20\%$. To determine corrected flux from the primary star or a target, we follow the logic of equation 1 in \citet{rampalli_hot_2019} and substitute the dilution term (neighboring flux over total flux) with the contamination ratio\footnote{A similar logic was applied to transit depths \citep[e.g.,][]{cooke_single_2018}}. We find that for a neighboring star varying by 5\%, its contribution to the light curve is 0.83\%. While the flux contribution decreases for dimmer neighboring stars. For example, a neighboring star 2 magnitudes dimmer will only contribute 0.13\%.

Our final sample comprises 30 M dwarfs that are well-separated from nearby stars in TESS: 27 M dwarfs that are likely single, 2 binaries with dim companions and 1 wide binary (Table~\ref{tab:per}). 

\subsection{Stellar Parameters}\label{subsec:param}

\subsubsection{Stellar Radii}
\label{subsec:rad}

There are multiple methods to calculate stellar radii: long baseline optical interferometry, analysis the orbit of low mass eclipsing binaries, or comparing spectra to stellar models. However the latter has been shown to underestimate the radius by around 4\%. \citep[e.g.,][]{kraus_mass-radius-rotation_2011,feiden_reevaluating_2012,spada_radius_2013,morrell_exploring_2019}. This is thought to be a result of magnetic activity which can inflate active stars relative to inactive stars. \citet[equation 4]{mann_how_2015} derive an empirical relationship between absolute $K$ magnitude ($M_K$) and the radius based on stars with interferometric radii: 
\begin{align}
R_* &= 1.9515 -0.3520M_K + 0.01680M_K^2
\end{align}
where $R_*$ is the stellar radius in $R_\odot$ and $M_K$ is absolute K magnitude.
We use apparent $K$ magnitude from the Two-Micron All-Sky Survey \citep[2MASS;][]{skrutskie_two_2006} and parallaxes from {\it Gaia} DR2  and DR3 catalog \citep[] []{gaia_collaboration_gaia_2016,gaia_collaboration_gaia_2018,gaia_collaboration_gaia_2022} to calculate $M_K$ and then $R_*$. We use Monte Carlo methods to propagate parallax errors through the calculation, which we add in quadrature with the rms scatter from the empirical relation: 2.89\%. The parallax errors ($3-4\%$) dominate the final errors on stellar radius. We show the radii measurements in Table~\ref{tab:per}

\begin{longrotatetable}
\begin{deluxetable*}{LCCCCCCCCCCCCC}
\centering
\tablecolumns{9}
\tablecaption{Parameters for some M dwarfs in Our Sample}
\tabletypesize{\footnotesize}
\tablehead{\colhead{TID} &  \colhead{2MASS} &  \colhead{$P_{rot}$ (d)}&  \colhead{Lit $P_{rot}$ (d)} & \colhead{Ref. $^a$} & \colhead{R$_*$ ($R_\odot$)} & \colhead{M$_*$ ($M_\odot$)} & \colhead{$v\sin i$ (km/s)} & \colhead{Ref.$^b$} & \colhead{Inclination ($\degree$)} &  \colhead{Rvar}}\
\startdata
3664898^{e} & 08294949+2646348 & 0.46$_${-1.12e-04}$^${+1.12e-04} &      0.46 & N16 & 0.124$_${-0.0037}$^${+0.0037} &  0.1 $\pm$ 0.002 &  9.6 $\pm$ 1.5 & R18 & 48.1 $\pm$ 10.9 & 0.006$_${-6.070e-05}$^${+6.150e-05} \\
 13960751 & 00544803+2731035 &  1.7$_${-5.12e-03}$^${+5.57e-03} &      1.70 & N16 & 0.282$_${-0.0084}$^${+0.0085} & 0.26 $\pm$ 0.006 &  9.0 $\pm$ 0.7 & K18 &  80.6 $\pm$ 7.4 & 0.006$_${-8.510e-05}$^${+8.350e-05} \\
17970570$^{c,d}$ & 11224274+3755484 & 0.56$_${-4.03e-04}$^${+4.08e-04} &      0.36 & N16 &  0.14$_${-0.0042}$^${+0.0042} & 0.11 $\pm$ 0.003 & 13.3 $\pm$ 0.8 & K18 &  81.6 $\pm$ 6.5 & 0.013$_${-1.871e-04}$^${+1.862e-04} \\
 53255031 & 05595569+5834155 & 0.95$_${-1.56e-03}$^${+1.49e-03} &      0.95 & N16 & 0.254$_${-0.0076}$^${+0.0076} & 0.22 $\pm$ 0.005 &  9.2 $\pm$ 1.7 & K18 & 47.2 $\pm$ 12.4 & 0.004$_${-3.510e-05}$^${+3.530e-05} \\
 53307637 & 07464203+5726534 & 0.82$_${-1.01e-03}$^${+1.03e-03} &      0.82 & N16 & 0.311$_${-0.0093}$^${+0.0093} & 0.29 $\pm$ 0.007 & 17.6 $\pm$ 0.8 & K18 &  69.1 $\pm$ 7.6 & 0.018$_${-6.990e-05}$^${+6.990e-05} \\
 58100576 & 00243478+3002295 & 1.08$_${-2.56e-03}$^${+2.56e-03} &      1.08 & N16 &  0.26$_${-0.0078}$^${+0.0078} & 0.23 $\pm$ 0.006 & 13.0 $\pm$ 0.8 & F18 &  81.9 $\pm$ 6.4 &  0.01$_${-7.080e-05}$^${+7.110e-05} \\
80859893$^c$ & 08012112+5624042 & 0.12$_${-1.15e-05}$^${+1.18e-05} &      0.12 & N16 & 0.143$_${-0.0042}$^${+0.0042} & 0.11 $\pm$ 0.003 & 66.0 $\pm$ 0.5 & K18 &  88.8 $\pm$ 1.1 & 0.022$_${-2.725e-04}$^${+2.914e-04} \\
 84649454 & 04121693+6443560 &  1.6$_${-1.02e-02}$^${+1.02e-02} &      1.59 & N16 & 0.221$_${-0.0065}$^${+0.0065} & 0.19 $\pm$ 0.004 &  7.2 $\pm$ 0.2 & K18 &  84.7 $\pm$ 4.2 & 0.003$_${-3.450e-05}$^${+3.620e-05} \\
85334035$^{c,e}$ & 11115176+3332111 & 7.74$_${-1.10e-01}$^${+1.15e-01} &      7.77 & H11 & 0.326$_${-0.0098}$^${+0.0098} & 0.31 $\pm$ 0.007 &  4.6 $\pm$ 0.7 & F18 &  78.9 $\pm$ 9.7 & 0.019$_${-3.420e-05}$^${+3.410e-05} \\
88099283$^c$ & 09535523+2056460 & 0.61$_${-7.62e-04}$^${+7.48e-04} &      0.62 & N16 & 0.179$_${-0.0053}$^${+0.0053} & 0.15 $\pm$ 0.003 & 10.1 $\pm$ 0.6 & K18 &  43.8 $\pm$ 3.3 & 0.003$_${-3.870e-05}$^${+4.100e-05} \\
155113423 & 23025250+4338157 & 0.35$_${-4.72e-04}$^${+4.73e-04} &      0.35 & H11 & 0.219$_${-0.0065}$^${+0.0065} & 0.19 $\pm$ 0.004 & 29.0 $\pm$ 1.5 & K18 &  67.9 $\pm$ 8.0 & 0.008$_${-5.820e-05}$^${+6.080e-05} \\
156491690$^{c,e}$ & 14311348+7526423 & 0.63$_${-1.10e-04}$^${+1.08e-04} &      0.63 & N16 &   0.199$_${-0.006}$^${+0.006} & 0.17 $\pm$ 0.004 & 14.3 $\pm$ 0.4 & K18 &  64.5 $\pm$ 3.5 & 0.006$_${-4.100e-05}$^${+3.960e-05} \\
173004464$^c$ & 10030191+3433197 & 0.86$_${-8.27e-04}$^${+8.24e-04} &      0.86 & N16 &   0.303$_${-0.009}$^${+0.009} & 0.28 $\pm$ 0.007 & 11.9 $\pm$ 0.2 & K18 &  41.8 $\pm$ 0.9 & 0.013$_${-7.160e-05}$^${+7.010e-05} \\
178947176$^{c,e}$ & 04353618-2527347 & 2.77$_${-6.05e-02}$^${+5.06e-02} &      2.78 & K12 & 0.403$_${-0.0122}$^${+0.0122} &   0.4 $\pm$ 0.01 &  7.1 $\pm$ 1.0 & F18 & 72.4 $\pm$ 11.8 & 0.003$_${-2.330e-05}$^${+2.210e-05} \\
187092382$^c$ & 03571999+4107426 & 0.57$_${-5.32e-04}$^${+6.32e-04} &      0.57 & N16 &   0.135$_${-0.004}$^${+0.004} & 0.11 $\pm$ 0.003 &  6.5 $\pm$ 0.8 & K18 &  33.5 $\pm$ 4.6 & 0.014$_${-1.877e-04}$^${+1.896e-04} \\
219463771 & 13533877+7737083 & 1.23$_${-5.33e-03}$^${+5.78e-03} &      1.23 & N16 &  0.29$_${-0.0087}$^${+0.0087} & 0.26 $\pm$ 0.006 &  8.9 $\pm$ 1.5 & R18 & 53.8 $\pm$ 13.5 & 0.008$_${-3.110e-05}$^${+3.110e-05} \\
229614158 & 18315610+7730367 & 0.86$_${-3.33e-04}$^${+3.26e-04} &      0.86 & N16 &   0.268$_${-0.008}$^${+0.008} & 0.24 $\pm$ 0.006 & 15.8 $\pm$ 0.7 & F18 &  81.2 $\pm$ 6.3 & 0.007$_${-6.850e-05}$^${+7.000e-05} \\
233068870 & 18021660+6415445 & 0.28$_${-3.32e-05}$^${+3.20e-05} &      0.28 & N16 & 0.195$_${-0.0058}$^${+0.0058} & 0.16 $\pm$ 0.004 & 10.3 $\pm$ 1.5 & R18 &  17.4 $\pm$ 2.5 & 0.003$_${-2.480e-05}$^${+2.400e-05} \\
252110114 & 01015952+5410577 & 0.28$_${-8.92e-05}$^${+8.81e-05} &      0.28 & N16 & 0.165$_${-0.0049}$^${+0.0049} & 0.13 $\pm$ 0.003 & 31.9 $\pm$ 3.1 & R18 &  79.0 $\pm$ 8.5 & 0.005$_${-8.340e-05}$^${+8.360e-05} \\
266744225$^c$ & 07444018+0333089 & 2.77$_${-2.48e-02}$^${+2.45e-02} &      2.80 & A98 & 0.333$_${-0.0099}$^${+0.0099} & 0.32 $\pm$ 0.007 &  4.0 $\pm$ 1.5 & R18 & 53.2 $\pm$ 18.5 & 0.007$_${-1.620e-05}$^${+1.700e-05} \\
273589987 & 19510930+4628598 & 0.59$_${-1.17e-03}$^${+1.34e-03} &      0.59 & H11 &  0.269$_${-0.008}$^${+0.0079} & 0.24 $\pm$ 0.006 & 22.9 $\pm$ 2.3 & R18 &  76.6 $\pm$ 9.7 & 0.007$_${-4.390e-05}$^${+4.030e-05} \\
283410775 & 22011310+2818248 & 0.45$_${-1.82e-04}$^${+1.68e-04} &      0.45 & M08 & 0.313$_${-0.0094}$^${+0.0094} & 0.29 $\pm$ 0.007 & 35.4 $\pm$ 3.5 & R18 &  76.5 $\pm$ 9.5 & 0.011$_${-3.860e-05}$^${+4.000e-05} \\
283866910$^e$ & 04171852+0849220 & 0.37$_${-2.05e-04}$^${+1.99e-04} &      0.37 & M22 & 0.271$_${-0.0081}$^${+0.0082} & 0.24 $\pm$ 0.006 & 37.3 $\pm$ 1.1 & K18 &  83.3 $\pm$ 4.9 &  0.01$_${-4.270e-05}$^${+4.020e-05} \\
286899254 & 15040626+4858538 & 1.02$_${-2.88e-03}$^${+2.79e-03} &      1.02 & H11 & 0.327$_${-0.0098}$^${+0.0098} & 0.31 $\pm$ 0.007 & 11.3 $\pm$ 2.0 & F18 & 48.8 $\pm$ 12.6 & 0.014$_${-1.575e-04}$^${+1.527e-04} \\
289972535 & 07555396+8323049 & 1.11$_${-1.18e-03}$^${+1.15e-03} &      1.11 & N16 & 0.274$_${-0.0083}$^${+0.0083} & 0.25 $\pm$ 0.006 & 13.4 $\pm$ 1.5 & R18 &  78.0 $\pm$ 9.2 & 0.019$_${-4.440e-05}$^${+4.550e-05} \\
291689873 & 23545147+3831363 & 4.75$_${-3.43e-02}$^${+3.33e-02} &      4.76 & K13 & 0.316$_${-0.0094}$^${+0.0095} &  0.3 $\pm$ 0.007 &  3.6 $\pm$ 1.5 & R18 & 66.8 $\pm$ 17.8 & 0.026$_${-6.350e-05}$^${+5.900e-05} \\
307913606 & 09245082+3041373 & 0.37$_${-5.18e-05}$^${+5.30e-05} &      0.37 & N16 & 0.421$_${-0.0125}$^${+0.0125} &  0.42 $\pm$ 0.01 & 44.9 $\pm$ 3.1 & K18 &  52.9 $\pm$ 5.6 & 0.027$_${-6.160e-05}$^${+6.720e-05} \\
309850275 & 15163731+5355457 & 0.52$_${-2.30e-04}$^${+2.65e-04} &      0.52 & N16 & 0.202$_${-0.0061}$^${+0.0061} & 0.17 $\pm$ 0.004 & 19.2 $\pm$ 0.4 & K18 &  81.6 $\pm$ 4.9 & 0.014$_${-1.357e-04}$^${+1.372e-04} \\
415508270$^e$ & 05041476+1103238 & 0.84$_${-2.11e-03}$^${+2.29e-03} &      0.84 & N16 &  0.19$_${-0.0059}$^${+0.0058} & 0.16 $\pm$ 0.004 & 10.8 $\pm$ 1.4 & K18 & 72.0 $\pm$ 11.5 & 0.003$_${-2.730e-05}$^${+2.660e-05} \\
440559522 & 06073185+4712266 & 0.86$_${-2.92e-03}$^${+3.00e-03} &      0.86 & N16 & 0.356$_${-0.0107}$^${+0.0107} & 0.34 $\pm$ 0.008 & 20.8 $\pm$ 0.8 & K18 &  81.0 $\pm$ 6.1 & 0.013$_${-7.090e-05}$^${+6.930e-05} \\
\enddata
\label{tab:per}
\tablecomments{
$^a$ Reference for literature periods: A98: \citet{alekseev_spottedness_1998}, M08: \citet{morin_large-scale_2008}, H11: \citet{hartman_photometric_2011}, K12: \citet{kiraga_asas_2012}, K13: \citet{kiraga_asas_2013}, N16: \citet{newton_rotation_2016}, and M22: \citet{medina_variability_2022}.  
$^b$ Reference for $v\sin i$: 
F18: \citet{fouque_spirou_2018}, 
R18: \citet{reiners_carmenes_2018}, and 
K18: \citet{kesseli_magnetic_2018}. 
$^c$ Needed manual flare removal on the edges after \texttt{stella}. $^d$ Needed jitter term toggled on. 
$^e$ Needed both the jitter and the trend term toggled on. 
A portion of the machine-readable table, which includes additional columns such as mean $H\alpha$ EWs, mean $L_{H\alpha}$/$L_{bol}$, Spectral type, RUWE, TESS Contamination Ratio, TESS magnitude, apparent K magnitude, parallax, parallax error and Month observed.}
\end{deluxetable*}
\end{longrotatetable}

\subsubsection{Stellar Mass}
\label{subsec:mass}

Using $M_K$ as a proxy for luminosity, we derive the masses of our targets following \citet{mann_how_2019}.
This relation applies to M dwarfs $0.075 < M/M_\odot < 0.70$ which encapsulate all of our targets. We use the corresponding Python package, \texttt{M\_-M\_K-}\footnote{\url{https://github.com/awmann/M_-M_K-}}, to calculate posterior probability distributions of mass for all 30 stars. 
We take the median of the distribution as our mass value, and compute the error as the 68\% confidence interval.

\subsubsection{Inclination}
\label{subsec:inc}

The $v \sin i$ of a star is determined from the rotational broadening of the spectral lines.
To accurately determine the $v \sin i$, we require high resolution spectra with $R= (\Delta \lambda/\lambda) \sim 20,000 - 40,000$ \citep{newton_rotation_2016}. We use $v \sin i$ from the literature \citep{fouque_spirou_2018, reiners_carmenes_2018, kesseli_magnetic_2018} along with our $P_\textrm{rot}$ and $R_*$ values to determine the inclination and its error for each target. 

\begin{figure}[htp!]
    \centering
    \includegraphics[width=\columnwidth]{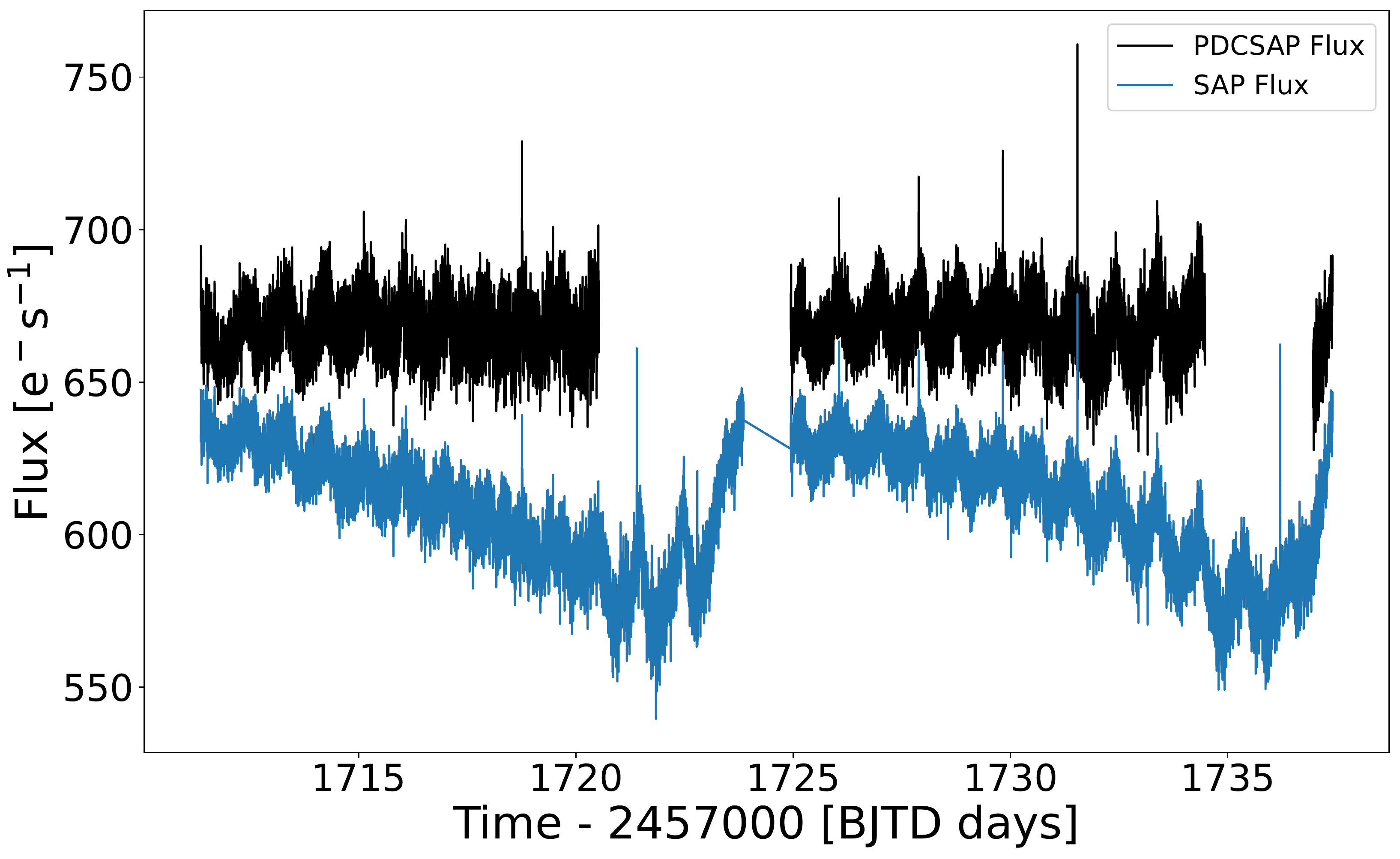}
    \caption{A \texttt{lightkurve} time-series comparing SAP flux and PDCSAP flux for TIC 343268431 (TESS magnitude 13.1). Long term systematic trends are seen from $\sim 1710$ to $\sim 1722$ Barycentric TESS Julian Date (BTJD - 2457000), the SAP flux decreases almost linearly and then increases after $\sim 1722$. We also see that the pattern repeats in the second observation cycle. Meanwhile, the PDCSAP flux remains constant aside from the rotation modulations.}
    \label{fig:pdc}
\end{figure}

We use  \texttt{inclinationmcmc}\footnote{\url{https://github.com/avanderburg/inclinationmcmc}} to calculate inclinations according to the framework from \citet{masuda_inference_2020}. The code calculates a posterior probability distribution for $\cos i$, where $i$ is the inclination. We then transform to $i$ assuming that $0<i<90\deg$. We adopt the median value as the best-fit $i$ and calculate error as the 68\% confidence interval.

\section{Observations and Data Reduction}\label{sec:obs}
\subsection{Optical Photometric Data}\label{subsec:tess}

TESS data products include 30-minute cadence full frame images and 2-minute cadence `postage stamp' images of selected stars \citep[$\sim200,000$;][]{ricker_transiting_2015}. The Science Processing Operations Center produces 2-minute cadence light curves FITS files which are then released to the public via Mikulski Archive for Space Telescopes (MAST) archive \citep{jenkins_tess_2016}.
We use the 2-minute cadence light curves.

Within each light curve file there are two different measurements of flux: the Simple Aperture Photometry flux (SAP) and the Pre-search Data Conditioned Simple Aperture Photometry (PDCSAP). SAP sums the photon counts of a stellar source and subtracts from it the photon counts of the sky background. SAP also removes cosmic rays and ``Argabrightening" events\footnote{non-systematic diffused light of unknown origins that affects the background and the source apertures} \citep{witteborn_debris_2011, JenkinsPA2020}. PDCSAP is derived from the SAP light curve, and accounts for systematics resulting from the spacecraft and its instruments, such as momentum dumps \citep{stumpe_kepler_2012, smith_kepler_2012, stumpe_multiscale_2014,ThompsonProducts2016}. Example SAP and PDCSAP light curves are shown in Figure \ref{fig:pdc}.

PDCSAP flux is effective at removing long-term trends.
\citet{newton_tess_2022} find that PDCSAP could also effect the observed rotation signature in some sectors; comparison of the SAP and PDCSAP light curves for stars in our sample indicates this is not a concern for us. 
Thus, for this paper we use the PDCSAP flux.

We use \lightkurve\  \citep{lightkurve_collaboration_lightkurve_2018, geert_barentsen_keplergolightkurve_2020} to download the TESS data. We use the quality bit-mask set to \texttt{default}, which removes bad cadences with major quality issues. 

We select TESS data taken within a sector or 27 days of the date of our spectroscopic observing runs. For example, our 2019 October observing run corresponds to TESS sectors 16 and 17 (2019 September 11 - 2019 October 07 and 2019 October 07 - 2019 November 02, respectively). If targets were not observed by TESS in either of these two sectors, we considered neighboring sectors such as sector 18 (2019 November 02 - 2019 November 27). We examine long \rvar \ in \S\ref{subsec:rvar} and have notice minute changes in \rvar \ for neighboring sectors. For observations that encompass multiple sectors, we apply \lightkurve's \texttt{stitch} method to stitch multiple sectors into a single light curve. Similar ranges are also use for the other observation runs and can be found in Table~\ref{tab:obs}.

\begin{deluxetable*}{LCCCCC}
\centering
\tablecolumns{5}
\tablecaption{Observations}
\tablehead{\colhead{Month} &  \colhead{Range of Observation (UTC)} & \colhead{Range of Observation (JD)} & \colhead{TESS Sectors} &  \colhead{Total Objects*} & \colhead{Spectra per Target}}\
\startdata
\text{2019 October} & 2458762.664754 - 2458767.595015 &	2019/10/06 - 2019/10/11 & 15-17 & 11 & 1\\
\text{2020 January} & 2458851.833316 - 2458856.002801 & 2020/01/03 - 2020/01/07 & 19 \ \& \ 20 & 9 & 3\\
\text{2020 February} & 2458880.911557 - 2458895.841905&	2020/02/01 - 2020/02/16 & 21 \ \& \ 22 & 9  & 2-4\\
\text{2020 December} & 2459188.724752 - 2459197.900665 & 2020/12/05 - 2020/12/14 & 32 & 4 & 25-135\\
\text{2021 January} & 2459222.855154 - 2459222.8949 &	2021/01/08 - 2021/01/08 & 33 & 1 & 68\\
\enddata
\label{tab:obs}
\tablecomments{*Note: There are 30 M dwarfs in total, but 3 of them were also observed a second time and 1 is the binary mentioned in Figure~\ref{fig:havarb}.}
\end{deluxetable*}

\subsection{Optical Spectroscopic Data}
\label{subsec:mdm24}

We observe $H\alpha$ emission using OSMOS \citep{martini_ohio_2011}, mounted on the 2.4m Hiltner telescope at MDM Observatory in Arizona. Our setup consists of the blue VPH grism with a inner 1.2" slit and a 4x1k region of interest. The resulting spectra span 390-680 nm, with peak efficiency around 640 nm and a $R\sim 1600$ resolution. In this configuration, OSMOS is well-suited to single-object spectroscopy, particularly for our study of H-alpha emission at around 656.21 nm.

\begin{figure}[htp!]
    \centering
    \includegraphics[width=\columnwidth]{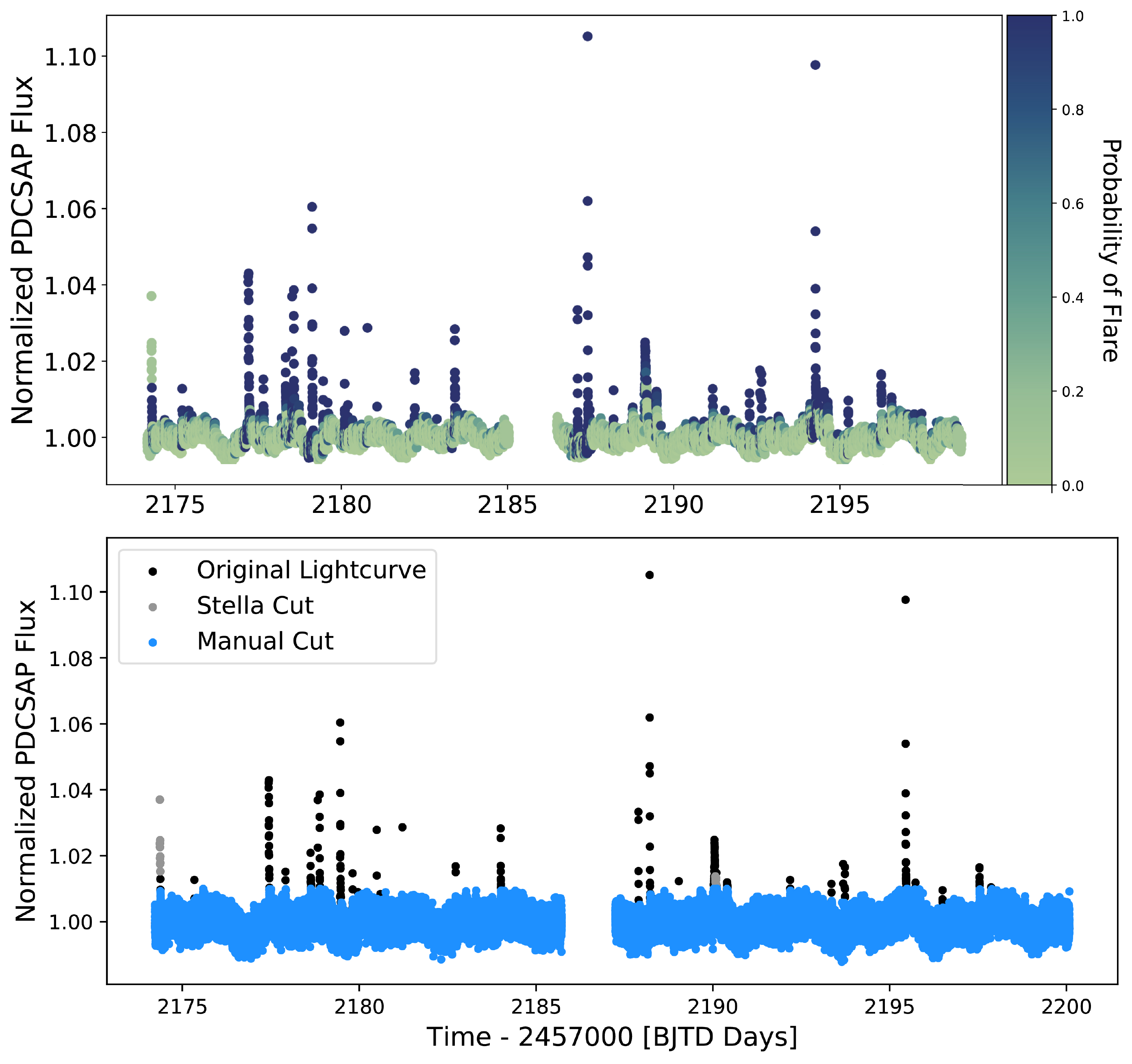}
    \caption{Normalized PDCSAP Flux vs.\ TESS Barycentric Julian Date (BJTD - 2457000). Top: The color bar represents the probability that each data point represents a flare, as determined by \stella. The probabilities range from 0 to 1, where one constitutes a flare. While most large flares were removed, the one in the beginning of this observation was only reduced in strength, so that the large flare around  2175~d needed to removed manually. All of the stars in the sample have more than one visible flare, while a handful have many flares such as this one. Bottom: Two of the flare removal methods discussed in \S\ref{subsec:flarerem}. This bottom plot was created for all of the stars in our sample and is available as supplemental material.}
    \label{fig:stella}
\end{figure}

Our spectroscopic observations took place during 5 queue runs between 2019 October and 2021 January (Table~\ref{tab:obs}). In total, we have 14 nights of data from OSMOS with the number of spectra increasing per observation cycle. For instance in 2019 October, we have 1 spectrum per star while in 2020 December we have more than 24 spectra per object. While a more specific account of the number of spectra can be found in the machine readable version of Table~\ref{tab:per}. The V magnitude of our stars vary from 17.47 to 11.19, which correlates to exposure times from 90s to 1200 seconds, respectively.

We use the \texttt{osmosreduct} pipeline provided by John Thorstensen\footnote{\url{https://github.com/jrthorstensen/thorosmos}}. With this semi-automated process, we perform standard flatfielding, bias subtraction, and wavelength calibration. We hand-check each file's aperture and remove bad points from the fit spectrum before extracting the 1D spectrum. 

For the wavelength calibration, we use a master wavelength calibration which was created using a median-combination of an HgNe + Ne lamp and a Xe lamp. We wavelength-bin the data by comparing the wavelengths of known emission features to their observed wavelengths in each file. We linearly fit the residuals and find the corresponding scale factor and zero point of the data. We use these two factors to adjust the master and subsequently bin the data. We anchor the fit to lines at 404.7, 435.8, 546.1, and 640.2 nm by-hand; the rest are fit automatically using a polynomial of order 6. 

The full spectra reduction amounts to for 63 objects with spectra range from 1-135 (Table~\ref{tab:obs}). TESS cuts, as discussed above, reduce our target list to 30 stars.

\section{Analysis}\label{sec:analysis}

\subsection{Light Curve Analysis}\label{subsec:flarerem}

We download \textit{TESS} PDCSAP data using \lightkurve \ for sectors coincident with our ground-based observations, and one sector before/after. 
As every star in our sample has at least one visible flare in the light curve, we first remove the flares (\S\ref{subsec:stella}). We then fit the rotation modulation (\S\ref{sec:gp-per}) and calculate the semi-amplitude (\S\ref{subsec:rvar}).

\subsubsection{Flare Removal}\label{subsec:stella}
We use \stella, a python package including a Convolutional Neural Network  \citep[CNN;][]{feinstein_flare_2020} that has been trained to identify flares. We utilize 100 \stella \ models trained on 2-min TESS data to detect the flares. Some models detect certain features better than others, so a combination of models yields more robust flare detections. 

The ensemble of models gives a score between 0 and 1 at each data point, where 0 is a non-flare and 1 is a flare. Although it is not a probability, a comparison of visually-identified flares and the score shows that its an acceptable estimator of a flare. We identify that anything with a score above 0.5 as a flare; this value was chosen based on visual inspection of light curve plots. This value is also used in \citet{feinstein_flare_2020}.

Some of the visual inspections show a few large flares that are missed by \stella \ (Figure~\ref{fig:stella}). Most of those flares are located in the beginning or end of the observation, or are lower energy but still apparent by eye. These may artificially elevate the semi-amplitude; thus, for 9 M dwarfs (flagged in Table~\ref{tab:per}), some features were removed manually by trimming the corresponding time stamps. 

Further outlier rejection is performed as part of our stellar rotation fitting, which is effective at removing the remaining flares or distorted flares.

\subsubsection{Gaussian Processes Fitting}\label{sec:gp-per} 

Prior to fitting for stellar variability, we bin the data from 2 minute cadence to 10 minutes using \lightkurve's \texttt{bin} function to increase computational efficiency. Rotational variability is on timescales of hours so this is not expected to impact our modeling.
 
A sinusoidal model is not always effective at modeling the rotational variability of M dwarfs, which can be quasi-periodic and non-sinusoidal. For this reason, we use Gaussian processes to fit the rotational variability \cite[e.g.,][]{angus_inferring_2018}. We modify the \starspot \ package to analyze the light curves \citep{angus_ruthangusstarspot_2021,https://doi.org/10.5281/zenodo.7697238}\footnote{\url{https://github.com/agarciasoto18/starrotate}}. 
The Gaussian process kernel function is composed of two stochastically-driven simple harmonic oscillators, which indicate the primary (the rotation period, $P_\textrm{rot}$) and secondary ($P_\textrm{rot}/2$) modes. 

Additional terms describe the amplitudes of variability, and how damped the two oscillators are. To account for other variability in the light curve, we can optionally include a jitter term that defines the white noise in the data and a trend term that accounts for long-term data unrelated to star spots. These two terms were only included when the code initially failed to converge on the correct period; otherwise, they have little effect on the fitting. We fit for jitter and/or long-term trends in 7 light curves (flagged in Table~\ref{tab:per}). 

Some of our uniform priors are uninformative. Once we are satisfied that we have selected the correct period of the star, we place a truncated normal prior of 30\% on $\log{P_{rot}}$ centered on the fiducial period of the star. This prevents unnecessary exploration of parameter space near $P_{rot}/2$.

First, we perform a maximum a posteriori (MAP) fit using \starspot, then create a prediction of the stellar rotation model. Then we calculate the residuals of the MAP prediction by subtracting the model from the data. Data is kept when the absolute value of the residual is smaller than 4 times the root-mean-square (rms) of the residuals. This code is run a total of three timea, rejecting outliers after each iteration.

Lastly, we sample the posterior of the stellar rotation model using the \pymc \ package within our modified version of \starspot \ \citep{salvatier_probabilistic_2016}. \starspot \ uses a Markov Chain Monte Carlo (MCMC) to find the best fit parameters. 

\begin{figure}
    \centering
    \includegraphics[width=\columnwidth]{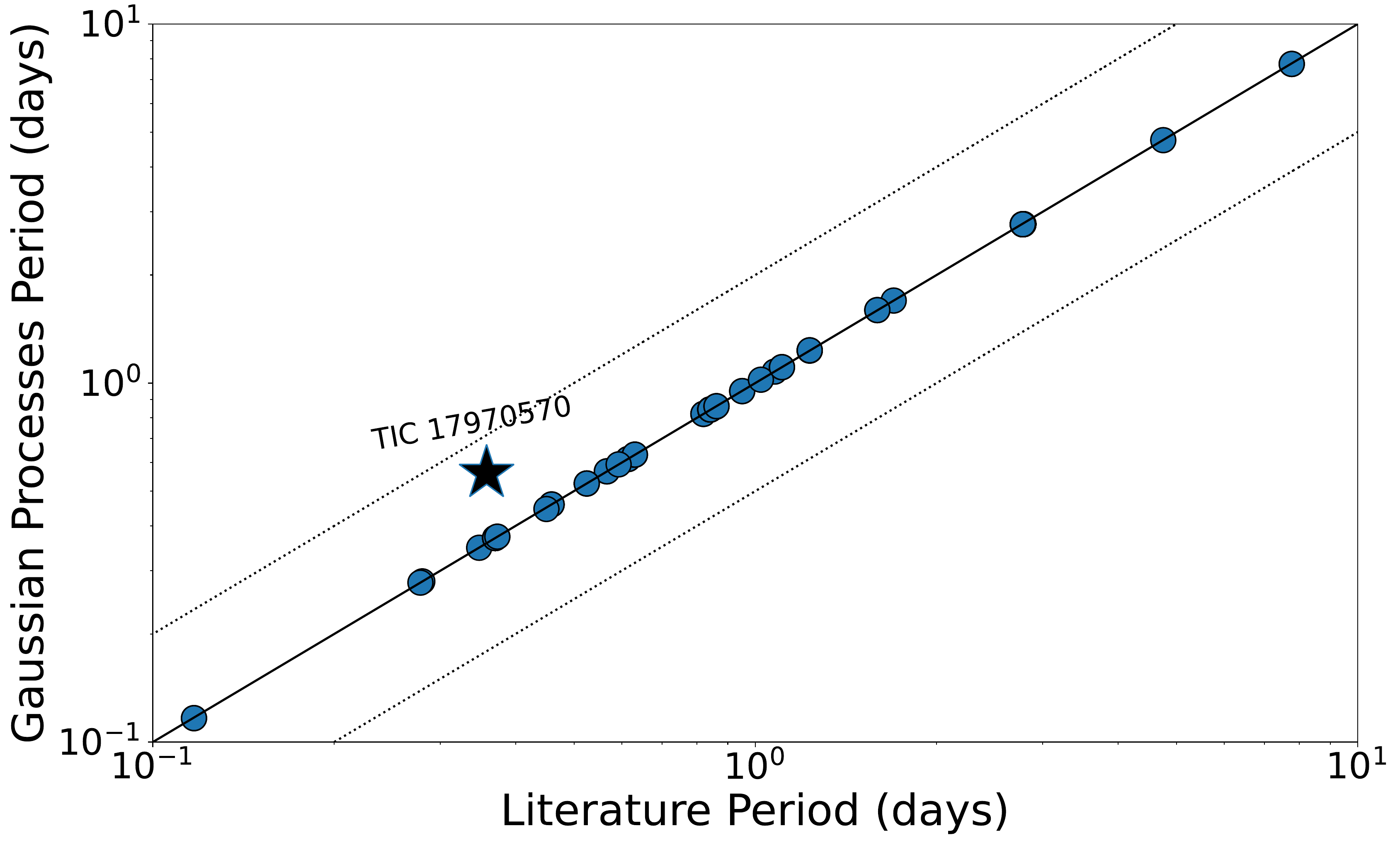}
    \caption{A comparison between the periods we find from \starspot \ and rotation periods in the literature. The black star symbol denotes TIC 17970570, where our period is not quite double the period from \citet{newton_rotation_2016}. Re-analysis of data from MEarth shows that $P_\textrm{rot}=0.5593\pm0.0004$ d is the true period, consistent with the result from TESS. We are therefore confident in our $P_\textrm{rot}$ values for all 30 of our targets.}
    \label{fig:newp}
\end{figure}

As detailed in Table~\ref{tab:per} and Figure~\ref{fig:newp}, the periods of 29 M dwarfs match well with the literature. However, for one M dwarf TIC 17970570 (star symbol), \starspot \ flagged a different period, of $0.5593\pm0.0004$ d. This is 1.6 times the literature value \citep[$0.358$ d;][]{newton_rotation_2016}. However, re-analysis of the data from MEarth's Data Release 11 \citep[2011-2012;][]{berta_transit_2012} data agrees with the period of 0.559 days, and we identify the 0.358 d signal as a 1 day alias. Thus, we can confirm the rotation periods for all 30 M dwarfs.

\subsubsection{Semi-Amplitude using \texorpdfstring{\rvar} \ Method}\label{subsec:rvar}

The $R_{var}$ method of calculating the photometric amplitude was first defined by \citet{basri_photometric_2010, basri_photometric_2011} and modified by \citet{mcquillan_statistics_2012, mcquillan_measuring_2013}.  We calculate $R_{var}$ by computing the difference between the 5th and 95th percentile of the median-normalized flux. $R_{var}$ uncertainties are calculated using a simple Monte Carlo method, in which we re-draw new flux values at random using a normal distribution centered on the original flux values with width set by the flux errors. We then calculate \rvar \ from the new fluxes. We perform 5000 iterations and pick the final $R_{var}$ to be the 50th percentile, with the uncertainties from the 16th and 84th percentiles.

\begin{figure}[htp!]
    \centering
    \includegraphics[width=\columnwidth]{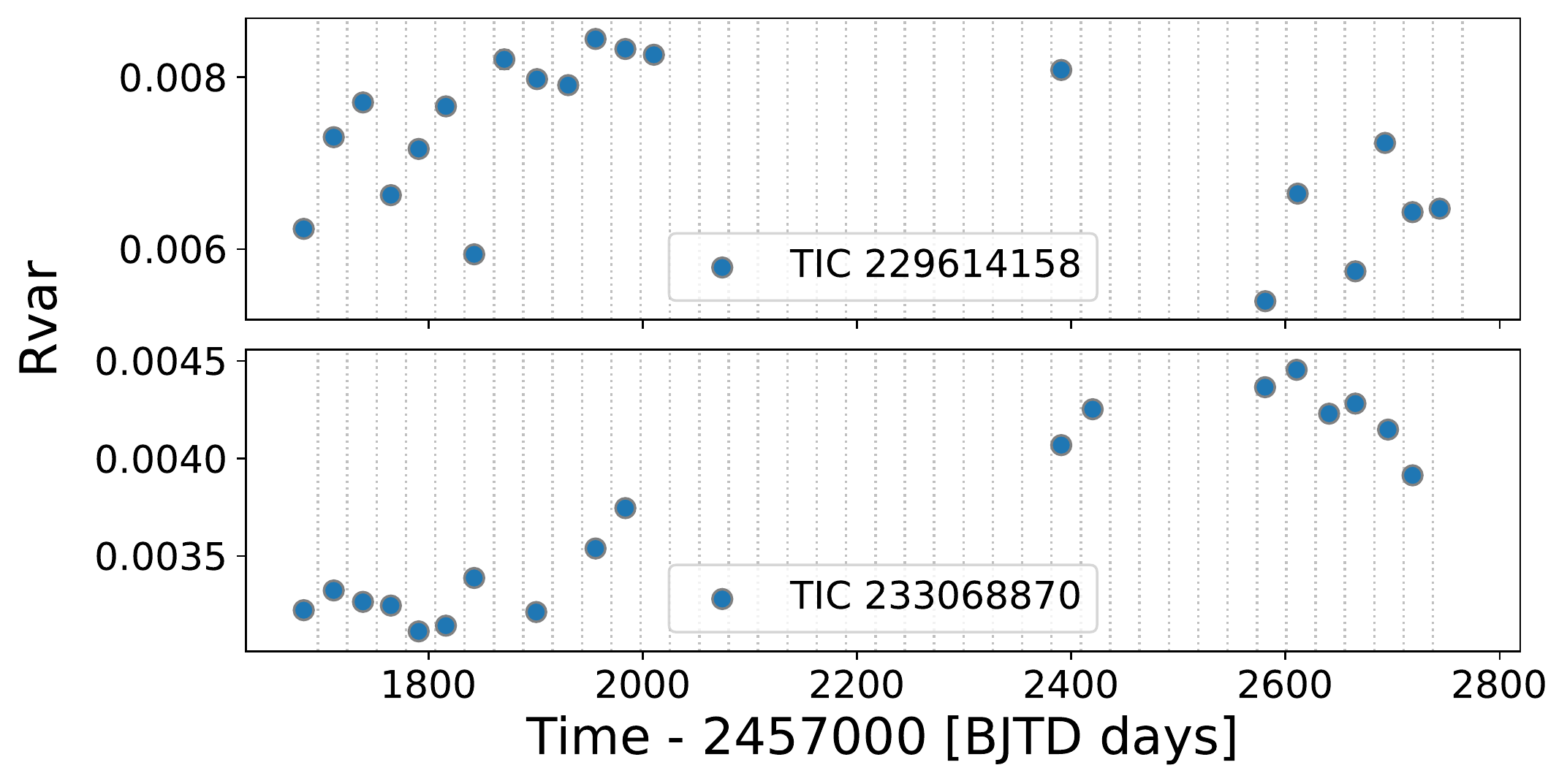}
    \caption{Two graphs of semi amplitude-\rvar \ plotted over time. Top: TIC 229614158 and Bottom: TIC 233068870. The vertical lines represent the length of a TESS sector, approximately every 27.4 days.}
    \label{fig:longr}
    
\end{figure}

We examine \rvar \ to set our definition of contemporaneous spectroscopic and photometric data, as we only have truly simultaneous observations for 18 out of our 30 objects. We pick stars in our sample with 5 or more sectors of TESS data, calculate \rvar \ for each sector, and plot the \rvar \ over time. In Figure \ref{fig:longr}, we show these plots for the two stars in our sample with the most TESS data: TIC 233068870 and TIC 229614158. For TIC 233068870, we remove one of the sectors from Figure \ref{fig:longr} due systematic trends/non-stellar variability, which produces an artificially large value for \rvar \ in that sector. Generally, \rvar \ for both stars remains constant for 2-3 sectors, but it seems that the morphology can remain relatively consistent up around 6 sectors. It's worth noting how  \rvar \ for both stars changes from one year to the next. This is also seen in \citet{robertson_persistent_2020}. For example, TIC 233068870 increases from $R_{var}\lesssim 0.0035$ to $R_{var}\sim 0.0045$ in the span of a year. Given the result, we use TESS data taken within one to two sectors of the observation date of our spectroscopic data.

Finally, it is important to consider that \rvar \ is sensitive to instrumental noise which has similar strengths to stellar variability, as with TIC 233068870 above. This would also mean toggling the long-term trend in \starspot \ might also affect the \rvar \ measurement. Although the \rvar \ method is more robust than fitting for the amplitude, the addition of noise increases the value of \rvar \ \citep{basri_comparison_2013, johnson_forward_2021}.

\subsection{Spectral Analysis}
\subsubsection{Calculating \texorpdfstring{\ha} \ Equivalent Widths}\label{subsec:ha}

To calculate $H\alpha$ Equivalent Widths (EWs), we first shift the spectra to rest wavelengths. To determine the velocity offset, we fit the $H\alpha$ lines in the M dwarf spectra with Gaussian distributions using \astropy's \texttt{specutils} \citep{astropy_collaboration_astropy_2013, the_astropy_collaboration_astropy_2018,nicholas_earl_astropyspecutils_2019}.

We use the standard definition of EW, following \citet{west_sloan_2011} and \citet{newton_h_2017} and summing partial pixels assuming they are uniformly illuminated.

\begin{align}
\textnormal{EW} = \sum \left(1-\frac{F(\lambda)}{F_c}\right)\delta\lambda,   \label{eqn:EW}
\end{align}
where $F(\lambda)$ is the flux, $\delta \lambda$ is the pixel size, and $F_c$ is the mean flux of the \ha continuum level regions 6500-6550 \AA \ and 6575-6625 \AA.  The summation is over width of the $H\alpha$ line, which is defined to be 8 \AA\ (6558.8 - 6566.8 \AA). We report $H\alpha$ EWs as negative values for emission.

\subsubsection{Removing the Basal \texorpdfstring{\ha} \ Contribution}

Following the methodology in \citet{newton_h_2017} we remove the absorption component of the $H\alpha$ emission line. In M dwarfs with relatively little chromospheric heating, the $H\alpha$ line is in absorption. As chromospheric heating increases the absorption line strength increases, reaching a maximum positive EW of $\sim 0.7$ \AA. To evaluate only the emission component, we must therefore define and account for the base absorption level at a given mass. 
As in \citet{newton_h_2017}, 
our spectral resolution is too low to distinguish between strengthening absorption or emission for a weakly active star. Thus, we only set the base emission level.


\citet{newton_h_2017} showed that as mass decreases, the maximum $H\alpha$ absorption EWs also decreases. 
We define a basal absorption value ($EW_{basal}$), and subtract it from the measured value ($EW_{measured}$, from Equation~\ref{eqn:EW}) to calculate the final relative value ($EW$).
\begin{align}
EW_{basal} &= -5.36570831 + 7.62481452\ M \nonumber \\
& - 2.88931216\ M^2 + 0.52686116\ M^3  \label{eqn:ewfit}\\
EW &= EW_{measured} - EW_{basal}   \label{eqn:ewrelative}
\end{align}
where $M$ is the stellar mass in $M_\odot$. We use these relative EWs for all further calculations including \lhalbol.

\begin{figure}[htp!]
    \centering
    \includegraphics[width=\columnwidth]{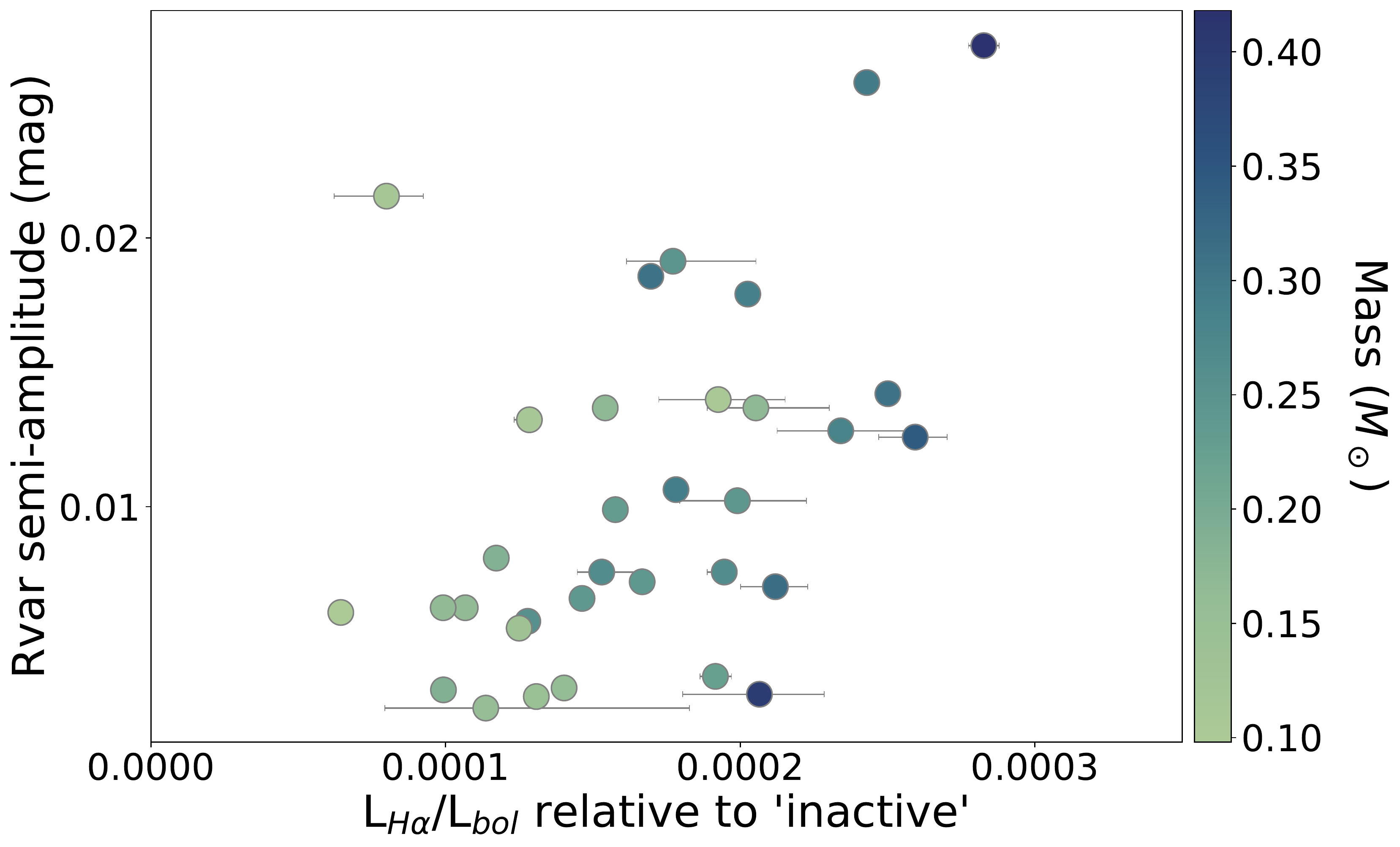}
    \caption{Semi amplitude, \rvar \, plotted against the $H\alpha$ strength, \lhalbol. The x-axis error bars for the OSMOS data in this case represent the range in \lhalbol. There is a possible correlation between \rvar\ and  \lhalbol, with more active stars showing higher photometric variability.}
    \label{fig:complha}
    
\end{figure}

We then normalize by the bolometric luminosity of the star to get \lhalbol. In practice, to calculate \lhalbol, we multiply $EW$ by the $\chi$ factor. This allows us to calculate \lhalbol \ without absolute flux calibrated data \citep{walkowicz__2004}. The $\chi$ factor is determined from the division of the spectral continuum flux (the mean flux between 6550-6560 and 6570-6580 \AA) by the apparent bolometric flux for stars with flux-calibrated data. The $\chi$ factor is calibrated using stellar models, then is calculated from color indices. 
We adopt the $\chi$ values from \citet{newton_h_2017}, which are derived following \citet{douglas_factory_2014}. 

\section{Results}\label{sec:results}
\subsection{\texorpdfstring{\ha} \ Activity vs Amplitude Relation}\label{subsec:actamp}

\citet{newton_h_2017} find that stars with greater $H\alpha$ activity have more photometric variability. As in that work, we plot the \lhalbol \ and amplitude (or in our case, \rvar; Figure~\ref{fig:complha}). Note that the \lhalbol \ error bars correspond to the range in \lhalbol \ values due to multiple spectroscopic observations. For example, the star with the largest range, TIC 415508270, has 135 measurements. In other words, some of these stars have $H\alpha$ emission variability (see \S\ref{subsec:havarb}). 

\begin{figure}[t!]
    \centering
    \includegraphics[width=\columnwidth]{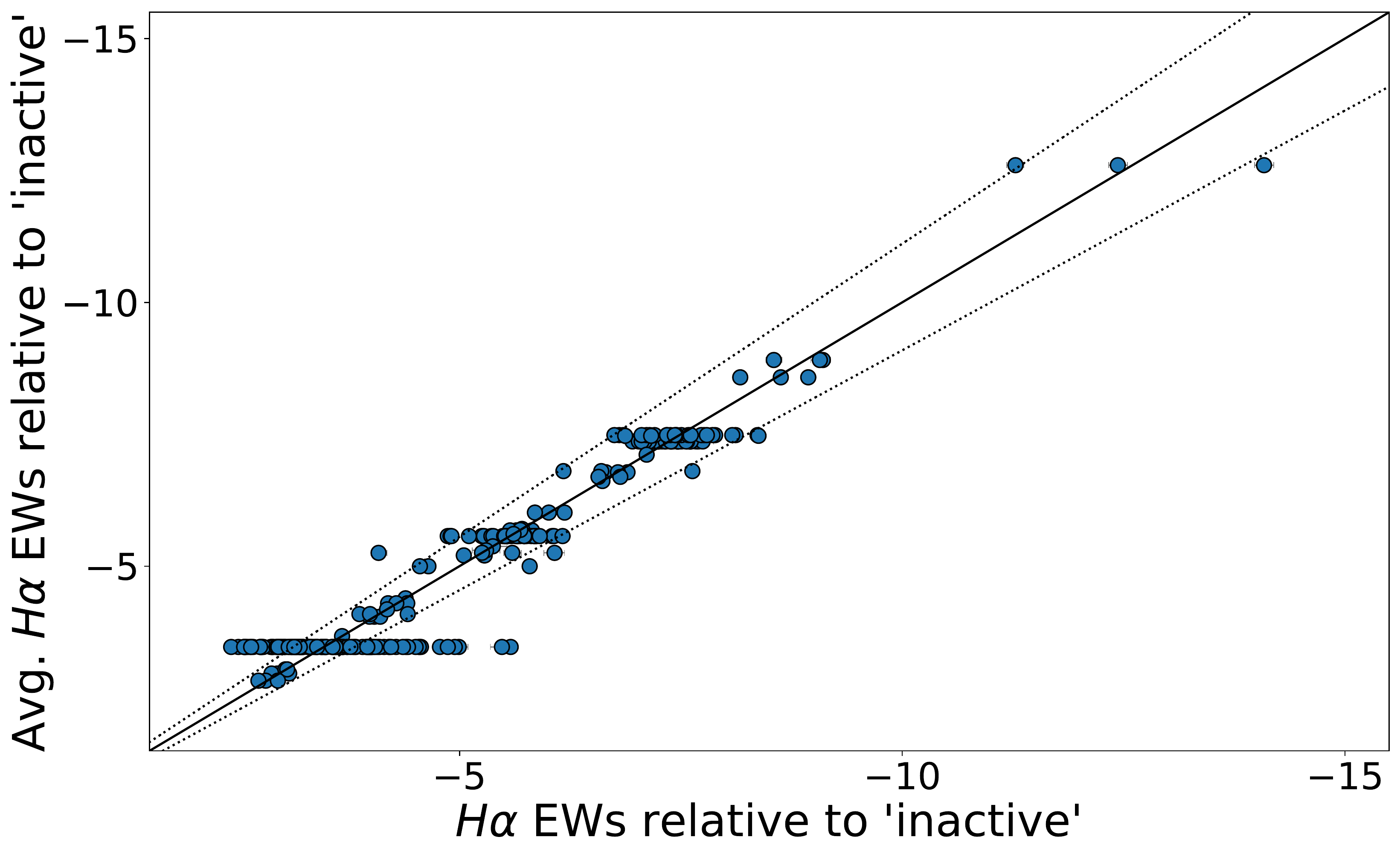}
    \caption{The comparison of the average relative $H\alpha$ EWs for each M Dwarf against their range of relative $H\alpha$ EWs. The dotted lines represent 10\% variation from the mean value. TIC 415508270 has the largest range: 30.2\% to 60.8\% variation from the mean value.}
    \label{fig:compha}
\end{figure}

We find a tentative positive correlation between \lhalbol \ and \rvar. The Spearman rank correlation ($\rho$) tests for a correlation between variables. Similar to how we extract the \rvar \ and its errors, we run a simple Monte Carlo methods program to randomize the $\rho$ and the $p$-values and set the low and high error bars to the 16th and 84th percentile. We calculate a $\rho = 0.479_{-0.169}^{+0.155}$ with $p=0.005_{-0.005}^{+0.075}$. This falls between 0 (poor correlation) and 1 (positive monotonic correlation), so we consider it a moderate positive correlation. Conversely, given the large error in $p$-value, we cannot say that this result is not by chance. 
In contrast, \citet{newton_h_2017} find evidence for a correlation with $\rho = 0.39\pm0.03$ and $p = 4 \times 10^{-8}$; further analysis is warranted.  

Figure~\ref{fig:complha} also shows the mass dependence of the relative \lhalbol \ \citep{newton_h_2017}. As the stellar mass increases, relative \lhalbol \ also increases.
We get $\rho = 0.755_{-0.115}^{+0.084}$ with 
$p= (0.01_{-0.01}^{+1.39})\times 10^{-4}$, 
meaning that the mass is positively correlated to relative \lhalbol. 

This is consistent with other studies, which find a decrease in \lhalbol \ for late-type M dwarfs \citep[$<$ M4 or M6; e.g.,][]{gizis_palomarmsu_2002,west_spectroscopic_2004,lee_short-term_2010, kruse_chromospheric_2010,bell_h_2012, zhang_stellar_2023}. 
This mass dependence is also seen in both the saturated and unsaturated regime with $\log R^\prime_{HK}$ \citep{schrijver_magnetic_1987,rauscher_ca_2006,houdebine_rotation-activity_2017,boudreaux_ca_2022}, but no such variation is seen in in X-rays \citep{wright_solar-type_2016}.
Finally, there is a mass-dependence in the saturated magnetic field strength for M dwarfs \citep{vidotto_stellar_2014}, which could perhaps explain the variations in $H\alpha$ activity.
This suggests that the magnetic field may be affecting the chromosphere differently than the corona. 

We conclude that contemporaneous spectroscopy and photometry, and a more uniform and flexible method of measuring semi-amplitudes, overall does not result in a clearer amplitude-activity relation. 
However, our data suggest that there may be a positive correlation between photometric semi-amplitude and $H\alpha$ emission in M dwarfs which is also observed in other spectra types \citep[][]{zhang_stellar_2023}.

\begin{figure*}[t]
    \centering
    \includegraphics[width=2.1\columnwidth]{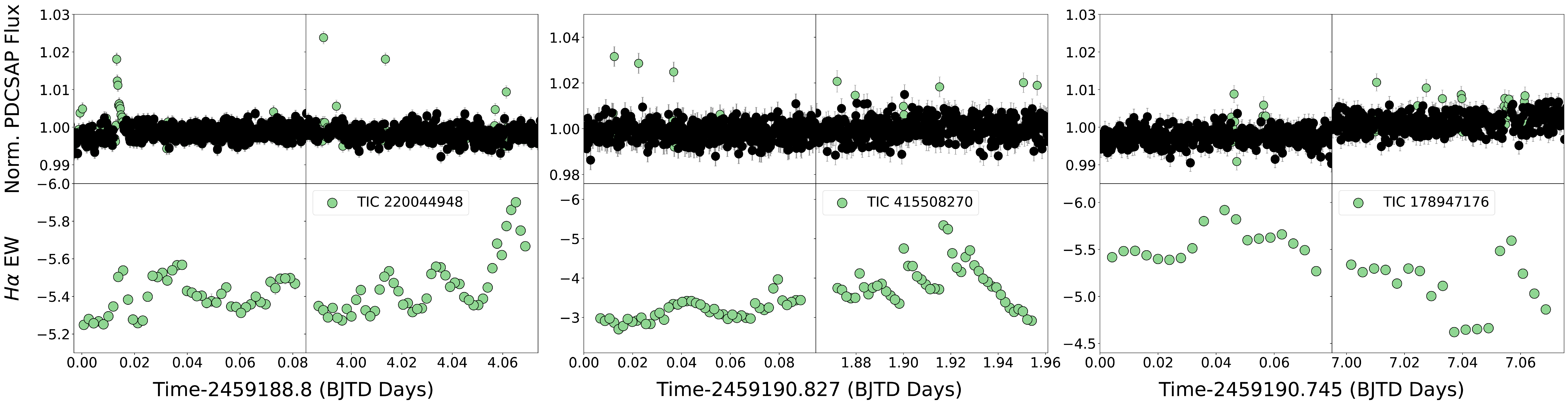}
    \caption{The original PDCSAP light curve on the top, $H\alpha$ EWs in the middle, and number of $\sigma_{std}$ from the mean $H\alpha$ EWs in the bottom against time. Left: 2 nights for TIC 220044948, Middle: 2 nights for TIC 415508270, Right:
    2 nights for TIC 178947176. The light green circles are the raw light curve and the black circles is the light curve after \stella\  removed the flares. For the middle and bottom panel, the light green circles are each target-specific $H\alpha$ EW observation.}
    \label{fig:havarb}
\end{figure*}

\subsection{\texorpdfstring{\ha} \ Variability}\label{subsec:havarb}

Here we investigate the variability of H$\alpha$ EWs for stars in our sample with multiple observations. Figure \ref{fig:compha} compares the average $H\alpha$ EWs to all $H\alpha$ EWs measured for that star. All stars with multiple spectra show some short term $H\alpha$ variability. Excluding stars with a single spectrum (11 stars), 7/19 stars show variability above 10\% of the mean. However, it is important to note that 16 stars in this subset have less than 4 spectra. The stars with more than 25 spectra, excluding a binary plotted in Figure~\ref{fig:havarb}, all show short-term variability. This result does not seem likely to be caused by sampling bias, but we need follow-up observations to confirm.

Next, we search for simultaneous $H\alpha$ and photometric variations. Flares can enhance $H\alpha$ emission, and \citet[][]{medina_variability_2022} showed that length of time for a flare are around 20-45 minutes which aligns with some $H\alpha$ variability timescales. Thus, 
in Figure Figure~\ref{fig:havarb}, we plot the TESS light curve on top and the $H\alpha$ EWs measurements on the bottom to see if $H\alpha$ emission variability can be attributed to flaring. Both our photometric and spectroscopic data sets align in time for three out of four objects observed on December 2020. All three show $H\alpha$ EWs varying on timescales of less than an hour, similar to \citep{medina_variability_2022}. 

TIC 220044948 (left) is an M2 star with $P_{rot} = 0.6$ days\---it has a RUWE $\sim 2$ so it was not included in the overall sample. On the first night of observations, it shows $H\alpha$ emission enhancement corresponding to a flare observed in TESS\---around 0.01 on the x-axis. 
However there are several peaks following that event, which could suggest that $H\alpha$ emission is from stochastic flaring. Alternative explanations for the variability include low energy flares below the capabilities of TESS  \citep{medina_variability_2022}, or contribution of a binary companion.

On the other hand, TIC 415508270 (middle) is not a binary. It is a fully convective M5 with a $P_{rot} = 0.84$ days. Like, TIC 220044948, it shows multiple $H\alpha$ EW peaks on its second night of observation. The rough time between each peak for TIC 220044948 and TIC 415508270 both happen on the order of 20-45 minutes. Although, in this case there isn't clear flare enhancement in the light curve. There are no clear flares in the light curve that align with the large $H\alpha$ peaks. 

Lastly, TIC 178947176 (right) is a fully convective M3.5 with $P_{rot} = 2.77$ days. It shows a single peak in the first night. Unlike the other two stars, this star does not have consecutive peaks so we cannot give it a rough timing estimate.

Overall, we see flare-induced enhancement in $H\alpha$ EW for one star, but the driving factor for other $H\alpha$ variability is still unclear. Another possibility is that the variability not tied to a flare or binary companion could be intrinsic in nature. Nevertheless, a more in depth analysis for these stars and a larger sample will be the subject of a future paper.

\section{Discussion: possible origins of scatter in the amplitude-activity relation}\label{sec:discussion}

Like \citet{newton_h_2017}, we find a positive relationship between \lhalbol \ and variability amplitude. 
\citet{newton_h_2017} find $\rho_{H\alpha} = 0.39 \pm 0.03$ with $p = 4 \times 10^{-8}$, while we find $\rho_{H\alpha} = 0.479_{-0.169}^{+ 0.155}$.
However, we find only weak evidence against the null hypothesis ($p = 0.005_{-0.005}^{+0.075}$) and we cannot confirm if this correlation is due to random chance at this time. There is no indication that contemporaneous observations or more robust amplitude measurements decrease the scatter in this relationship. Hence, the scatter in $H\alpha$ emission might be an intrinsic property of M dwarfs, unrelated to their overall spot coverage. Below, we consider possible explanations for our results.

\subsection{Inclination of the stellar rotation axis}
\begin{figure}[!t]
    \centering
    \includegraphics[width=\columnwidth]{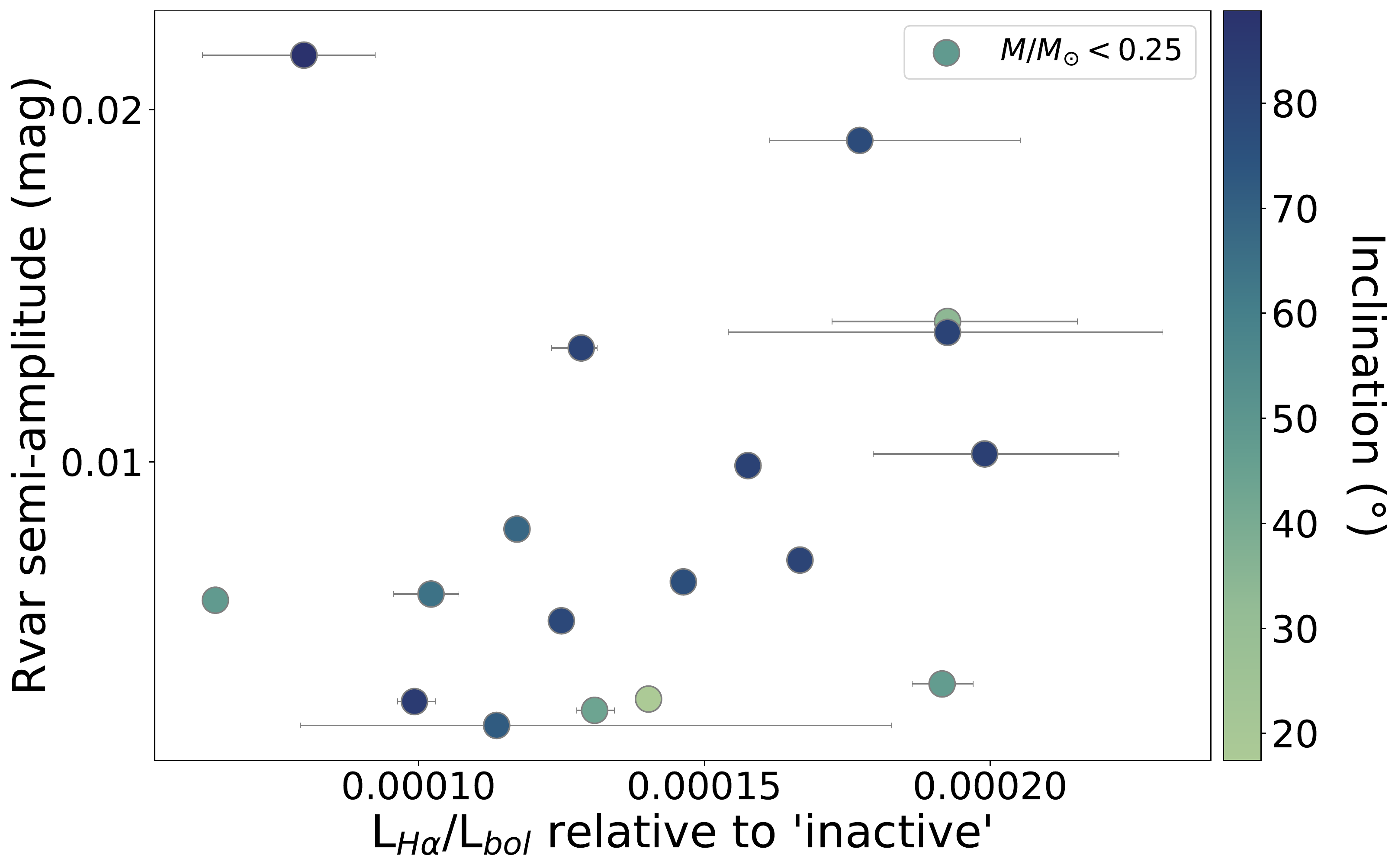}
    \includegraphics[width=\columnwidth]{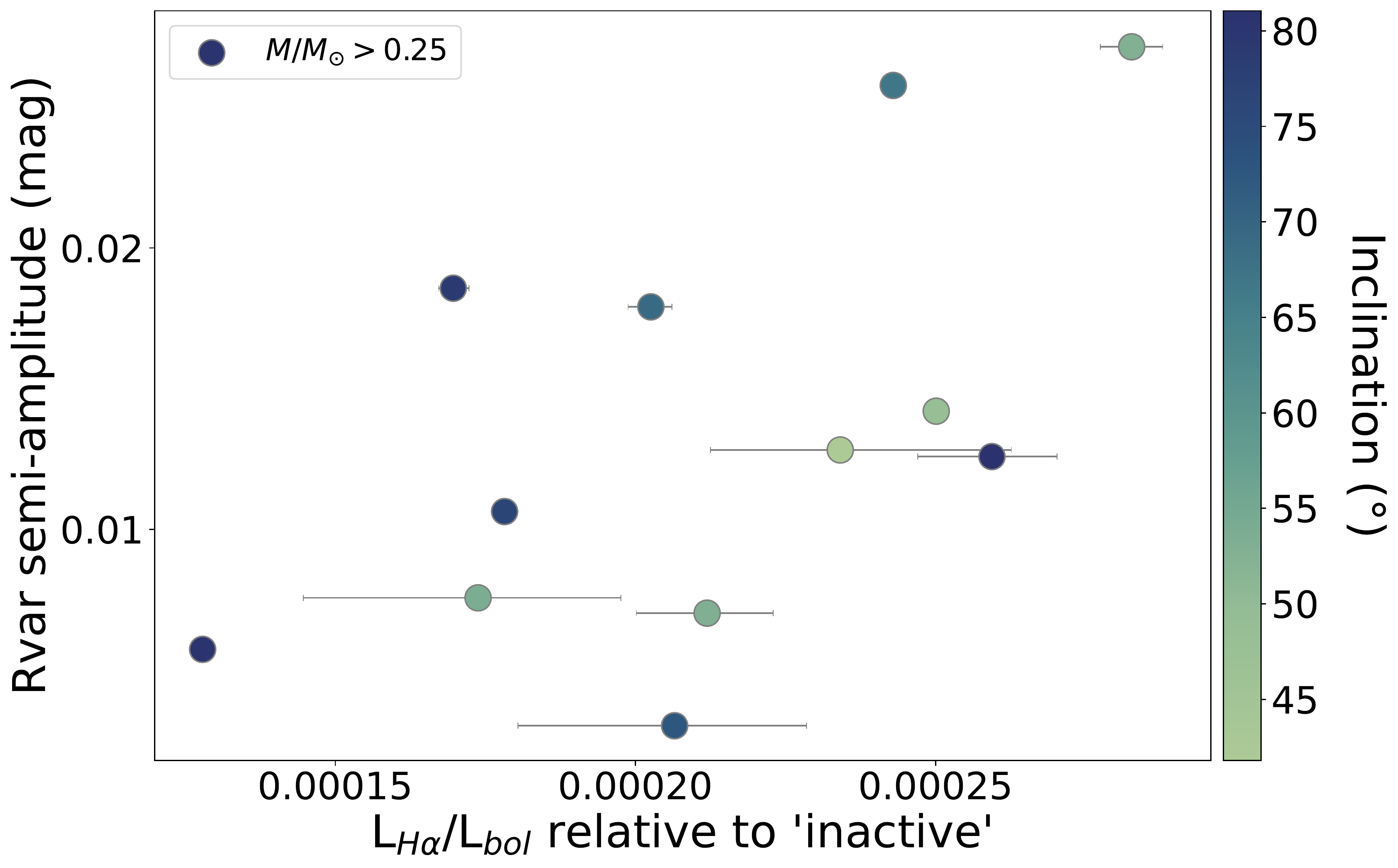}
    \caption{Same set up as Figure~\ref{fig:complha}, except the color bar is inclination. Top: Stars with mass $< 0.25$ $M_\odot$. Bottom: Stars with mass $> 0.25$ $M_\odot$, separated as detailed in \citet{newton_rotation_2016}. The ``error bars'' in this case represent the full range in $H\alpha$ luminosity measurements. Inclination does not visibly correlate with with L$_{H\alpha}$/L$_{bol}$ but there might some indication in the low-mass bin that smaller \rvar \ have smaller inclinations.}
    \label{fig:inc}
    
\end{figure}
On the Sun, and possibly some other solar-like stars, spot groups appear where the magnetic field rises through the stellar surface, and migrate  towards the equator as a cycle progresses \citep{vogt_doppler_1999,strassmeier_doppler_2000}. While for fast rotating cool dwarfs, it is theorized that the spots groups appear at poles due to the greater strength of the Coriolis force over the buoyancy force \citep{schuessler_why_1992}. Thus, many permutations of stellar inclination and spot latitude can produce the same photometric amplitudes \citep{basri_information_2020}. This is because \rvar \ is not tracing the spot coverage but rather the hemispheric spot distribution asymmetry, thus \rvar \ is also susceptible to spot evolution \citep[e.g.,][]{luger_starry_2019,basri_information_2020,luger_mapping_2021}. Moreover, low-amplitude variability could indicate large pole-on/polar spots that are almost always in view \citep{berdyugina_magnetic_2002,berdyugina_starspots_2005,rackham_transit_2019,luger_mapping_2021}, but could also be due to small spots in an equator-on view.

However, we find no correlation with inclination (Figure~\ref{fig:inc}). The Spearman rank correlation coefficients for both mass bins are not consistent with a correlation and the $p-$value indicates that it's likely due to random errors. For masses, $< 0.25\ M_\odot$, $\rho_{i} = 0.365_{-0.261}^{+0.230}$, $p =  0.135_{-0.126}^{+0.458}$, while for Mass $> 0.25\ M_\odot$: $\rho_{i} = -0.175_{-0.266}^{+0.315}$ and $p = 0.457_{-0.332}^{+0.372}$. We would expect that smaller \rvar \ would correspond to smaller inclinations (pole-on). While, the relationship between the lower mass bin and the inclination show a very weak positive correlation, accounting for the error bars, the $p-$ values are large. 

This lack of observed correlation may be due to our small sample size and lack of pole-on rotators. 
For instance, in Table~\ref{tab:per}, we see the lowest inclination to be $17.43 \pm 2.53$, while the highest is $88.95\pm0.93$ (equator-on). The majority (20/30) fall on the higher end of inclinations ($\gtrsim 65\degree$) thus we are lacking pole-on stars which might drive a more significant correlation.  Given these limitations, for our small sample of 30 M dwarfs, we cannot confirm if inclinations have a correlation with photometric amplitude. 
\begin{figure}[!t]
    \centering
    \includegraphics[width=\columnwidth]{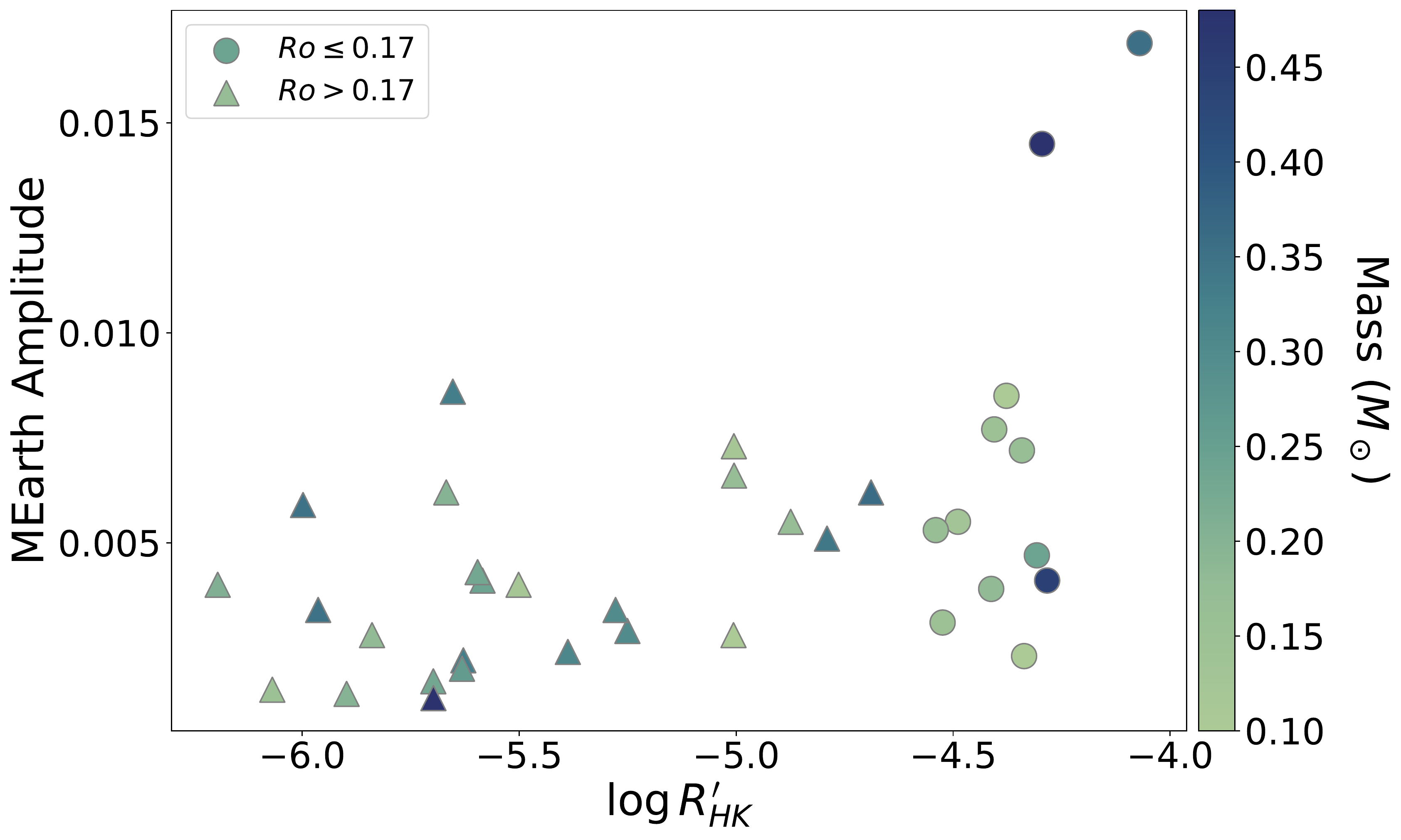}
    \caption{This plot is the MEarth fitted-amplitude plotted against the $\log R^\prime_{HK}$ (RUWE $< 1.6$) using data from \citet[][]{boudreaux_ca_2022}. In order to compare to our sample, we differentiate stars in the unsaturated regime (triangles) and stars in the saturated regime (circles). The stars in the saturated regime are defined as stars with Rossby $< 0.17 \pm 0.04$.}
    \label{fig:TB}
    
\end{figure}

\subsection{Activity and photometric variability as tracers of underlying stellar surface}

The scatter between \lhalbol \ and \rvar \ seems to be intrinsic. One possibility for the scatter is our choice of proxy for starspots: photometric amplitude. Amplitudes encode multiple degenerate spot properties such as spot distribution, coverage, temperature and size. In addition, we are ignoring contribution from bright spots such as plages and faculae. Albeit to a lesser extent, these bright regions can also impact the photometry \citep[][]{jackson_why_2012,jackson_relationship_2013,rackham_transit_2018,rackham_transit_2019,johnson_forward_2021}.

Although $H\alpha$ is the primary activity indicator for M dwarfs, several solar studies use Ca II H\&K ($\log R^\prime_{HK}$) as an chromospheric activity indicator \citep[e.g.,][]{karoff_observational_2016,mandal_association_2017,morris_are_2018,morris_stellar_2019}. We therefore test whether we observe any correlation between \rvar\ and $\log R^\prime_{HK}$ (Figure~ \ref{fig:TB}). 

We apply the selection criteria from our sample, including RUWE $< 1.6$, to the Ca II H\&K sample in \citet[][]{boudreaux_ca_2022}. An affect of our choice in instrument is the location of our stars in the saturated regime, thus we also cut by the critical Rossby number $<0.17\pm0.04$ listed in the paper. The relationship between amplitude and $\log(R^\prime_{HK})$ for stars in the saturated regime (circles) is $\rho_{HK} = 0.348_{-0.334}^{+0.285}$, $p = 0.246_{-0.219}^{+0.457}$, a weakly positive correlation that could be due to chance. Conversely, if we include stars in the unsaturated regime (triangles), or stars with periods longer than around 15 days, we get $\rho_{HK}=0.498_{-0.154}^{+0.132}$ and $p_{HK} = 0.002_{-0.002}^{+0.038}$ or a statistically significant positive relationship. We propose that including unsaturated stars may make the correlation clear enough to improve $p-$value as well. If this is the case, our $H\alpha$ results may also hinge on the inclusion of unsaturated stars.

Finally, while $H\alpha$ and Ca II H\&K are both tracers for plages, photometric amplitude traces starspots. 
There have been multiple studies that show that small plages can be co-located with spots \citep[e.g.,][]{radick_patterns_1998,radick_patterns_2018,jarvinen_ek_2007,jarvinen_magnetic_2008,morris_are_2018,fang_stellar_2020}. \citet{mandal_association_2017} compared plages and spots on the sun and found that on longer timescales, such as the 11 year solar-cycle, these two features are very correlated. On shorter timescales such as days and months, however, they are not correlated due to the shorter lifetime of spots. \citet{preminger_sunspot_2007} find a similar result for faculae and spots. While, we try to account for spot evolution by using contemporaneous data, it is therefore possible that spots and plagues are not well-correlated in M dwarfs, or that our results are affected by the degeneracies in spot properties derived from photometric amplitudes. 

\subsection{\texorpdfstring{\ha} \ Short-term Variability} \label{subsec:amber}

Lastly, a third possibility could be the short-term variability of $H\alpha$. Previous works have demonstrated significant short-timescale variability of $H\alpha$ \citep[e.g.,][]{kruse_chromospheric_2010, lee_short-term_2010,medina_variability_2022,duvvuri_fumes_2023} and Ca II H\&K \citep[e.g.,][]{suarez_mascareno_magnetic_2016,kumar_study_2023,duvvuri_fumes_2023}. For example, \citet{kruse_chromospheric_2010} shows that 74\% of their M dwarf sample (with $H\alpha$ emission and independent of spectral type) vary between 15 minutes to an hour while \citet{medina_variability_2022} finds variability within 20-45 minutes. We also find short term variation within that range with $H\alpha$ EWs peaks increasing 1-3 $\sigma_{std}$ over the mean value. 

The dominant source for the short-term variability might be low-energy flares \citep[e.g.,][]{liebert_flaring_2003,lee_short-term_2010,hilton_galactic_2011,medina_flare_2020,medina_variability_2022,zhang_stellar_2023,duvvuri_fumes_2023}. A flare occurs when a magnetic field loop breaks: part of it snaps off and releases energy while the other part reconnects at a lower energy state. Some of that released energy heats the chromosphere and enhances $H\alpha$ emission \citep{cram_model_1979,medina_flare_2020,medina_variability_2022}. 

\citet{medina_flare_2020} show fully convective M dwarfs show an activity-flare rate relation similar to the saturated rotation-activity relation: the logarithmic flaring rate per day decreases sharply when H$\alpha$ emission decreases below a critical value ($H\alpha$ EW $<-0.71$\AA). Interestingly, they find that in the ``saturated'' regime, the flare rate shallowly increases with increasing emission ($<-0.71$\AA) in the saturated regime. This could explain the flaring activity described in \S\ref{subsec:stella}. Although the authors note it was unclear if the individual EWs are enhanced by flares, \citet{medina_variability_2022} compare the timescales of low-energy flare decay to the $H\alpha$ emission variability and find them to be similar.

Moreover, \citet[][]{maehara_time-resolved_2021} look at contemporaneous photometric data and $H\alpha$ spectra for YZ CMi (TIC 266744225, amplitude $\sim 3.6\%$ in the TESS band). Previous inclination measurements places it between $36_{-14}^{+17}$
\citep{baroch_carmenes_2020} and $60$ degrees \citep[][]{morin_large-scale_2008}, which matches our results of $60 \pm 18$. 
During a epoch of low flaring activity, they find that large $H\alpha$ emission corresponded to dips in rotational modulation. However, for an epoch with high flaring activity there was no significant modulation in $H\alpha$ EW. The believe that the difference could be due a difference in inclination and spot latitudes.
This would suggest the short-term variability in $H\alpha$ is due to flaring from newly emerging small active regions, in agreement with \citet[][]{medina_variability_2022}. 

Hence, the dominate source of $H\alpha$ enhancement for these stars of high activity seems to be the low-energy flares rather than the active regions on the star. As we saw in \S\ref{subsec:havarb}, TIC 220044948 shows clear $H\alpha$ enhancement during it's first night of observation similar to the one seen in \citet{medina_variability_2022}. Alternatively, we also see similar peaks and a larger $H\alpha$ peak during the second night that do not align with large flares in TESS\--also true for the second night of TIC 415508270. Given the rough periodicity of the peaks, between 20-45 minutes, we believe these could be due to low energy flares that are below the sensibility of TESS detection. We also noticed only 7/19 stars with variability greater than 10\% of the $H\alpha$ EW mean. These stars all have more than one spectrum, but only 3 M dwarfs with more than 25 spectra. Thus, we would need more time-resolved $H\alpha$ spectra in order to match more flares with the optical photometry. 

\section{Summary}\label{sec:conclude}

We expand on the work of \citet{newton_h_2017} by obtaining contemporaneous $H\alpha$ spectroscopy and photometric data from TESS. 
Our goal is to tighten the relationship between photometric variability and $H\alpha$ emission. However, we find no significant change in the relationship between photometric amplitude and \lhalbol.  We suggest scatter is not explained by the lack of simultaneity, stellar mass, or inclination. Other sources of scatter must dominate such as low-energy flares. 

Consistent with previous works, we saw that the photometric variability amplitude could remain constant for multiple months. In contrast, we saw that $H\alpha$ emission varies on shorter timescales than rotation. \citet{medina_variability_2022} suggest that small flares are the dominate source of these short-term fluctuations corresponding to flare decays of 20-45 minutes. However, this does not mean spots do not also contribute. Further research into spot properties and activity in M dwarfs will elucidate the connection between these phenomena and the diversity amongst M dwarfs. 

As we saw with the Spearman correlation coefficient and the $p$-value for Ca II H\&K and amplitude relationship, when we included the stars in the unsaturated regime the relationship strengthened. Thus, it is possible that if we include stars the unsaturated regime, we could get a smaller $p$-value and a stronger overall relationship. However, the large $p$-value for both $H\alpha$ luminosity and Ca II H\&K with photometric amplitude measurements could be due attributed to small number statistics. It is possible that a larger sample of high-activity stars could identify a clearer trend.

\begin{acknowledgments}
We are really grateful to proxy observer Justin Rupert for taking the spectroscopic data. We also thank John R. Thorstensen, Bur\c{c}in Mutlu-Pakdil, Brian C. Chaboyer, Thomas M. Boudreaux, Keighley E. Rockcliffe, Rayna Rampalli, Andrew Vanderburg and Kathryn E. Weil for their contributions, advice and conversations that improved the manuscript.

This work has made use of data from the European Space Agency (ESA) mission
{\it Gaia} (\url{https://www.cosmos.esa.int/gaia}), processed by the {\it Gaia}
Data Processing and Analysis Consortium (DPAC,
\url{https://www.cosmos.esa.int/web/gaia/dpac/consortium}). Funding for the DPAC
has been provided by national institutions, in particular the institutions
participating in the {\it Gaia} Multilateral Agreement. This paper includes data collected by the {\it TESS} mission. Funding for the {\it TESS} mission is provided by the NASA's Science Mission Directorate.
\end{acknowledgments}

%

\facilities{MDM:Hiltner \citep{martini_ohio_2011}, TESS \citep{ricker_transiting_2015}}


\software{\astropy \ \citep{astropy_collaboration_astropy_2013,  the_astropy_collaboration_astropy_2018}, 
\celerite \  \citep{foreman-mackey_celerite_2017,dan_foreman-mackey_dfmcelerite_2020},
\texttt{edmcmc} \ \citep{vanderburg_avanderburgedmcmc_2021},
\ernlib \ (\url{https://github.com/ernewton/ernlib}),  
\texttt{exoplanet} \citep{dan_foreman-mackey_exoplanet-devexoplanet_2020,agol_analytic_2020},
\texttt{inclinationmcmc} (\url{https://github.com/avanderburg/inclinationmcmc}),
\IRAF \ \citep{tody_iraf_1986,tody_iraf_1993,national_optical_astronomy_observatories_iraf_1999},
\lightkurve \ \citep{lightkurve_collaboration_lightkurve_2018, geert_barentsen_keplergolightkurve_2020},  
\texttt{M\_-M\_K-} \citep[\url{https://github.com/awmann/M_-M_K-};][]{mann_how_2019},
\texttt{NumPy} \citep{oliphant_guide_2006}, 
\pymc \ \citep{salvatier_probabilistic_2016},
\readmultispec \ \citep{kevin_gullikson_general-scripts_v10_2014}, 
\texttt{SciPy} \citep{virtanen_scipy_2020}, 
\texttt{specutils} \citep{nicholas_earl_astropyspecutils_2019}, 
\starspot \ \citep{angus_ruthangusstarspot_2021}, 
\stella \ \citep{feinstein_stella_2020}, 
\thorosmos \ (\url{https://github.com/jrthorstensen/thorosmos}), 
\thorsky \ (\url{https://github.com/jrthorstensen/thorsky})}

\bibliography{references}{}
\bibliographystyle{aasjournal}



\end{document}